\documentclass[twocolumn,pra,aps,floatfix,longbibliography]{revtex4-1}
\usepackage{amsmath,esint}
\usepackage{amsfonts}
\usepackage{amssymb}
\usepackage{gensymb}
\usepackage{siunitx}
\usepackage{verbatim}
\usepackage{graphicx}
\usepackage{xfrac}
\usepackage{times}
\usepackage{natbib}
\usepackage{hyperref}
\usepackage{dsfont}
\usepackage{url}
\usepackage{siunitx}
\usepackage{appendix}
\begin{document}

\title{Non-local topological electromagnetic phases of matter}

\author{Todd Van Mechelen}
\author{Zubin Jacob}
\email{zjacob@purdue.edu}
\affiliation{Birck Nanotechnology Center and Purdue Quantum Center, Department of Electrical and Computer Engineering, Purdue University, West Lafayette, Indiana 47907, USA}

\begin{abstract}

In 2+1D, nonlocal topological electromagnetic phases are defined as atomic-scale media which host photonic monopoles in the bulk band structure and respect bosonic symmetries (e.g. time-reversal $\mathcal{T}^2=+1$). Additionally, they support topologically protected spin-1 edge states, which are fundamentally different than spin-\sfrac{1}{2} and pseudo-spin-\sfrac{1}{2} edge states arising in fermionic and pseudo-fermionic systems. The striking feature of the edge state is that all electric and magnetic field components vanish at the boundary - in stark contrast to analogs of Jackiw-Rebbi domain wall states. This surprising open boundary solution of Maxwell's equations, dubbed the quantum gyroelectric effect [Phys. Rev. A \textbf{98}, 023842 (2018)],  is the supersymmetric partner of the topological Dirac edge state where the spinor wave function completely vanishes at the boundary. The defining feature of such phases is the presence of temporal and spatial dispersion in conductivity (the linear response function). In this paper, we generalize these topological electromagnetic phases beyond the continuum approximation to the exact lattice field theory of a periodic atomic crystal. To accomplish this, we put forth the concept of microscopic photonic band structure of solids - analogous to the traditional theory of electronic band structure. Our definition of topological invariants utilizes optical Bloch modes and can be applied to naturally occurring crystalline materials. For the photon propagating within a periodic atomic crystal, our theory shows that besides the Chern invariant $\mathfrak{C}\in\mathbb{Z}$, there are also symmetry-protected topological (SPT) invariants $\nu\in\mathbb{Z}_N$ which are related to the cyclic point group $C_N$ of the crystal $\nu=\mathfrak{C}\mod N$. Due to the rotational symmetries of light $\mathcal{R}(2\pi)=+1$, these SPT phases are manifestly bosonic and behave very differently from their fermionic counterparts $\mathcal{R}(2\pi)=-1$ encountered in conventional condensed matter systems. Remarkably, the nontrivial bosonic phases $\nu\neq 0$ are determined entirely from rotational (spin-1) eigenvalues of the photon at high-symmetry points in the Brillouin zone. Our work accelerates progress towards the discovery of bosonic phases of matter where the electromagnetic field within an atomic crystal exhibits topological properties. 

\end{abstract}

\maketitle 

\section{Introduction}

From a material science standpoint, all known topological phases of matter to date have been characterized by electronic phenomena \cite{Kane2010,Ryu2010}. This is true for both time-reversal broken phases - often called Chern insulators \cite{Haldane1988,Zoller2003,Houck2010,Jotzu2014} and time-reversal unbroken phases - known as topological insulators \cite{Kane2005,Bernevig1757,Maciejko2011}. The signature of time-reversal broken phases is the quantum Hall conductivity $\sigma_{xy}=ne^2/h$, which is quantized in terms of the electronic Chern invariant $n\in\mathbb{Z}$ \cite{Klitzing1980,Laughlin1981,Thouless1982}. $e$ being the elementary charge of the electron and $h$ the Planck constant. Only recently has the idea of bosonic Hall conductivity and topological bosonic phases been put forth \cite{Wen2011,Chen1604,Vishwanath2012,Vishwanath2013,Metlitski2013,Senthil2013,Lin2015,Tian2016,Zhang2018}.

However, it should be emphasized that the traditional Hall conductivity \cite{Shimano2010,Zheng2002} only has topological significance, with respect to the electron, in the static $\omega=0$ and local $k=0$ limits of the electromagnetic field $\sigma_{xy}(0,0)=ne^2/h$. At high frequency $\omega\neq 0$ and short wavelength $k\neq 0$, the Hall conductivity $\sigma_{xy}(\omega,\mathbf{k})$ acquires new physical meaning. We have shown that the electromagnetic field itself becomes topological \cite{VanMechelen2018,van_mechelen_photonic_2018} and nonlocal Hall conductivity functions identically to a photonic mass \cite{Dunne1999,BOYANOVSKY1986483,Horsley2018} in the low-energy physics $\omega\approx 0$. These topological electromagnetic phases of matter depend on the \textit{global} behavior of $\sigma_{xy}(\omega,\mathbf{k})$, over all frequencies and wave vectors.

As of yet, only the continuum topological theory of the aforementioned quantum gyroelectric effect (QGEE) has been solved \cite{VanMechelen2018,van_mechelen_photonic_2018}. Our goal is to extend this concept beyond the long wavelength approximation to the exact lattice field theory of optical Bloch waves. In this regime, we must consider not only the first spatial component $\sigma_{xy}(\omega,\mathbf{k})=\sigma_{xy}(\omega,\mathbf{k},0)$ but all spatial harmonics of the crystal $\mathbf{g}\neq 0$, to infinite order,
\begin{equation}
 J_x^{\textrm{Hall}}(\omega,\mathbf{k})=\sum_\mathbf{g}\sigma_{xy}(\omega,\mathbf{k},\mathbf{g})E_y(\omega,\mathbf{k}+\mathbf{g}).
\end{equation}
$\mathbf{g}\cdot\mathbf{R}\in 2\pi\mathbb{Z}$ are the reciprocal lattice vectors and $\mathbf{R}$ is the primitive vector of the crystal. In this case, $E_i$ is the \textit{microscopic} electric field. The electromagnetic field must be described to the same scale as the electronic wave functions, i.e. for photon momenta on the order of the lattice constant $ka=\pi$, with $a\approx \SI{5}{\angstrom}$. Since topological invariants are fundamentally global properties, these astronomically deep subwavelength fields actually play a role in the topological physics.

The idea of lattice topologies in electromagnetism was first proposed by Haldane \cite{Haldane2008,Haldane2008_2} in the context of photonic crystals \cite{Wang2009,Wang:05,Lu2013,HafeziM.2013,Zheng2015,Karzig2015,Alu2017,Lein2017}. These are artificial materials composed of two or more different constituents which form a macroscopic crystalline structure. A few important examples are gyrotropic photonic crystals \cite{Lu2013,Wang2009,Wang:05}, Floquet topological insulators \cite{Lindner2011,Cayssol2013,Rechtsman2013} and bianisotropic metamaterials \cite{Khanikaev2017,Slobozhanyuk2017,He4924,GLYBOVSKI20161,Sihvola2005,Alaee:12,Papasimakis:10,Li2018,ShanuiFan2017,CTChan2016,Silveirinha2015,Gao2015,Jin2016,Hassani2017,Song2018}. Instead, we focus on the microscopic domain and utilize the periodicity of the atomic lattice itself. Thus, the topological invariants in our theory are connected to the microscopic atomic lattice and not artificially engineered macroscopic structures. We stress that in the microscopic case, the electromagnetic theory is manifestly bosonic \cite{BB1987,Stone2016,Stone1432,Gawhary2018} (e.g. time-reversal $\mathcal{T}^{2}=+1$) and characterizes topological phases of matter fundamentally distinct from known fermionic and pseudo-fermionic phases.

With that in mind, this paper is dedicated to solving two longstanding problems, which is of interest to both photonics and condensed matter physics. The first, is developing the rigorous theory of optical Bloch modes in natural crystal solids. This problem gained significant interest in the 60's and 70's in the context of spatial dispersion (nonlocality) as it lead to qualitatively new phenomena - such as natural optical activity (gyrotropy) \cite{Hubbard1955,Falk1960,Penn1962,landau2013electrodynamics}. The current paper builds on our recent discovery of the quantum gyroelectric effect \cite{VanMechelen2018,van_mechelen_photonic_2018} where we have shown that nonlocality is also essential for topological phenomena and is a necessary ingredient in any long wavelength theory. However, since topological field theories are global constructs, a complete picture can only be achieved in the microscopic domain of Bloch waves. Most of the foundations have been summarized by Agronovich and Ginzburg in their seminal monograph on crystal optics \cite{agranovich2013crystal}. Nevertheless, topological properties have never been tackled to date and a few fundamental quantities, such as the Bloch energy density, have not been defined.

This leads to the second problem - deriving the electromagnetic topological invariants of these systems given only the atomic lattice. We solve this problem and also provide a systematic bosonic classification of all 2+1D topological photonic matter. Utilizing the optical Bloch modes, we show that a Chern invariant $\mathfrak{C}\in\mathbb{Z}$ can be found for any two-dimensional crystal and characterizes distinct topological phases. We then go one step further and classify these topological phases with respect to the symmetry group of the crystal - the cyclic point groups $C_N$. These are known as symmetry-protected topological (SPT) phases \cite{Fu_Liang2011,Hughes2011,Slager2012,Fang2012,Kruthoff2017,Bradlyn2017,Song2017,Sedrakyan2017,Po2017,Thorngren2018,Matsugatani2018,Watanabet2018} and the spin of the photon is critical to their definition. The rotational symmetries of light $\mathcal{R}(2\pi)=+1$ impart an intrinsically bosonic nature to these phases, which are fundamentally different than their fermionic counterparts $\mathcal{R}(2\pi)=-1$ encountered in conventional condensed matter systems. We illustrate this fact by directly comparing SPT bosonic and fermionic phases side-by-side. Our rigorous formalism of microscopic photonic band structure provides an immediate parallel with the traditional theory of electronic band structure in crystal solids. 

This article is organized as follows. In Sec.~\ref{sec:Lattice_Photonics} we develop the general formalism of 2+1D lattice electromagnetism. First we derive the generalized linear response function - accounting for spatiotemporal dispersion to infinite order in the crystal's spatial harmonics $\mathbf{g}$. Thereafter, we find the equivalent Hamiltonian that governs all light-matter Bloch excitations of the material. In Sec.~\ref{sec:Rotational_Symmetry} we study the discrete rotational symmetries (point groups) of the crystal and the implications on spin-1 quantization \cite{Bradlynaaf5037,Zhu2017,Hu2018,Hu2018_2,Fulga2018,Tan2018} of the photon. The following Sec.~\ref{sec:Bosonic_Phases} discusses the electromagnetic Chern number and its relationship to symmetry-protected topological (SPT) bosonic phases. The bosonic classification of each phase is related directly to integer quantization of the photon [Tab.~\ref{tab:SPTPhases}] and this is compared alongside their fermionic counterparts [Tab.~\ref{tab:SPTPhasesFermion}]. Sec.~\ref{sec:Conclusions} presents our conclusions.

The focus of this paper is 2+1D topological electromagnetic (bosonic) phases of matter $\mathfrak{C}\neq 0$ which requires breaking time-reversal symmetry. These bosonic Chern insulators are ultimately related to nonlocal gyrotropic response (Hall conductivity) and show unidirectional, completely transverse electro-magnetic (TEM) edge states \cite{VanMechelen2018,van_mechelen_photonic_2018}. However, time-reversal symmetric topological phenomena can arise in higher dimensional systems in the context of nonlocal magnetoelectricity \cite{van_mechelen2017}. These time-reversal symmetric phases possess counter-propagating TEM edge states and are interpreted as two copies of a bosonic Chern insulator. Features of topological phenomena, such as spin-momentum locking \cite{Yin1405,VanMechelen:16,kalhor_universal_2016,Pendharker:18,Bliokh1448}, have also been reported in conventional surface state problems - surface plasmon-polaritons (SPPs), Dyakonov waves, etc. However, these traditional surface properties are not connected to any topologically protected edge states or nontrivial phases.

\paragraph*{\textbf{Note:}}
Due to the frequency of integral formulas, all differential elements are assumed to be correctly normalized, such as Fourier transforms $d\omega/\sqrt{2\pi}\to d\omega$ and Fourier series $e^{i\mathbf{g}\cdot\mathbf{r}}/\sqrt{V}\to e^{i\mathbf{g}\cdot\mathbf{r}}$, where $V$ is the unit cell area of a crystal.

\begin{table*}
\centering
\caption{Summary of 2+1D topological electromagnetic (bosonic) phases. Symmetry-protected topological (SPT) bosonic phases exist in all cyclic point groups $C_{N=2,3,4,6}$. The continuous group $C_\infty$ describes the long wavelength theory $k\approx 0$. The topological phases are characterized by their Chern invariant $\mathfrak{C}\in\mathbb{Z}$ and SPT invariant $\nu\in\mathbb{Z}_N$. These numbers are not independent - but intimately related by the symmetries of the crystal: $\nu=\mathfrak{C}\mod N$. $\nu$ is protected by $N$-fold rotational symmetry and determines the Chern number up to a factor of $N$. The bosonic classification of $\nu$ represents the direct product of rotational eigenvalues $(\eta_N)^N=+1$ (roots of unity) of the electromagnetic field at high-symmetry points (HSPs) in the Brillouin zone. For the spin-1 photon, this classification is more intuitively understood in terms of modulo integers $m_N\in \mathbb{Z}_N$, which determine the $N$ possible eigenvalues of $\eta_{N}=\exp\left[i\frac{2\pi}{N} m_N\right]$.}
\label{tab:SPTPhases}
\begin{tabular}{lllll}
\hline
Point group, $C_N$  & Symmetry, $\mathbb{Z}_N$ & Bosonic classification, $(\eta_N)^N=+1$ & Boson SPT invariant, $\nu=\mathfrak{C}\mod N$ \\
\hline
$C_1$              & -                          & -                                                  & -                             \\
$C_2$              & $\mathbb{Z}_2$             & $\exp\left(i2\pi\mathfrak{C}/2\right)= \eta_2(\Gamma)\eta_2(\textrm{X})\eta_2(\textrm{Y})\eta_2(\textrm{M})$                          & $\nu=m_2(\Gamma)+m_2(\textrm{X})+m_2(\textrm{Y})+m_2(\textrm{M}) \mod 2$  \\
$C_3$              & $\mathbb{Z}_3$             & $\exp\left(i2\pi\mathfrak{C}/3\right)= \eta_3(\Gamma)\eta_3(\textrm{K})\eta_3(\textrm{K}')$                         & $\nu=m_3(\Gamma)+m_3(\textrm{K})+m_3(\textrm{K}') \mod 3$       \\
$C_4$              & $\mathbb{Z}_4$             & $\exp\left(i2\pi\mathfrak{C}/4\right)= \eta_4(\Gamma)\eta_4(\textrm{M})\eta_2(\textrm{Y})$        & $\nu=m_4(\Gamma)+m_4(\textrm{M})+2m_2(\textrm{Y}) \mod 4$   \\
$C_6$              & $\mathbb{Z}_6$             &  $\exp\left(i2\pi\mathfrak{C}/6\right)= \eta_6(\Gamma)\eta_3(\textrm{K})\eta_2(\textrm{M})$ & $\nu=m_6(\Gamma)+2m_3(\textrm{K})+3m_2(\textrm{M}) \mod 6$\\
$C_\infty$         & $\mathbb{Z}$               & $\exp\left(i\theta \mathfrak{C}\right)=\eta_\theta(0)\eta^*_\theta(\infty),~~\eta_{\theta}=\exp(i\theta m)$                      & $\nu=\mathfrak{C}= m(0)-m(\infty)$\\   
\hline
\end{tabular}
\caption{Summary of 2+1D SPT fermionic phases for comparison. The fermionic classification of $\nu$ represents the direct product of rotational eigenvalues $(\zeta_N)^N=-1$ (roots of negative unity) of the spinor field at HSPs in the Brillouin zone. For the spin-\sfrac{1}{2} electron, this classification is more intuitively understood in terms of modulo half-integers $m_N\in\mathbb{Z}_N+\sfrac{1}{2}$, which determine the $N$ possible eigenvalues of $\zeta_{N}=\exp\left[i\frac{2\pi}{N} m_N\right]$.}
\label{tab:SPTPhasesFermion}
\begin{tabular}{lllll}
\hline
Point group, $C_N$  & Symmetry, $\mathbb{Z}_N$ & Fermionic classification, $(\zeta_N)^N=-1$ & Fermion SPT invariant, $\nu=\mathfrak{C}\mod N$ \\
\hline
$C_1$              & -                          & -                                                  & -                             \\
$C_2$              & $\mathbb{Z}_2$             & $\exp\left(i2\pi\mathfrak{C}/2\right)= \zeta_2(\Gamma)\zeta_2(\textrm{X})\zeta_2(\textrm{Y})\zeta_2(\textrm{M})$                          & $\nu=m_2(\Gamma)+m_2(\textrm{X})+m_2(\textrm{Y})+m_2(\textrm{M}) \mod 2$  \\
$C_3$              & $\mathbb{Z}_3$             & $\exp\left(i2\pi\mathfrak{C}/3\right)= -\zeta_3(\Gamma)\zeta_3(\textrm{K})\zeta_3(\textrm{K}')$                         & $\nu=m_3(\Gamma)+m_3(\textrm{K})+m_3(\textrm{K}')+\sfrac{3}{2} \mod 3$       \\
$C_4$              & $\mathbb{Z}_4$             & $\exp\left(i2\pi\mathfrak{C}/4\right)= -\zeta_4(\Gamma)\zeta_4(\textrm{M})\zeta_2(\textrm{Y})$        & $\nu=m_4(\Gamma)+m_4(\textrm{M})+2m_2(\textrm{Y})+2 \mod 4$   \\
$C_6$              & $\mathbb{Z}_6$             &  $\exp\left(i2\pi\mathfrak{C}/6\right)= -\zeta_6(\Gamma)\zeta_3(\textrm{K})\zeta_2(\textrm{M})$ & $\nu=m_6(\Gamma)+2m_3(\textrm{K})+3m_2(\textrm{M})+3 \mod 6$\\
$C_\infty$         & $\mathbb{Z}$               & $\exp\left(i\theta \mathfrak{C}\right)=\zeta_\theta(0)\zeta^*_\theta(\infty),~~\zeta_{\theta}=\exp(i\theta m)$                      & $\nu=\mathfrak{C}= m(0)-m(\infty)$\\   
\hline
\end{tabular}
\end{table*}

\section{Lattice electromagnetism}\label{sec:Lattice_Photonics}

\subsection{2+1D electrodynamics}

In this paper we focus on two-dimensional materials and the topological electromagnetic phases associated with them. The preliminaries for 2+1D electromagnetism can be found in Appendix A of Ref.~\cite{VanMechelen2018}. Conveniently, the restriction to 2D limits the degrees of freedom of both the electromagnetic field and the induced response of the material, such that strictly transverse-magnetic (TM) waves propagate. The corresponding wave equation reads,
\begin{equation}\label{eq:wave_equation}
\mathcal{H}_0f=i\partial_t g, \qquad f=\begin{bmatrix}
E_x\\ E_y \\ H_z
\end{bmatrix}, \qquad g=\begin{bmatrix}
D_x\\ D_y \\ B_z
\end{bmatrix}.
\end{equation}
$f$ is the TM polarization state of the electromagnetic field and the material response is captured by the displacement field $g$. $\mathcal{H}_0(\mathbf{p})=\mathbf{p}\cdot\mathbf{S}$ are the vacuum Maxwell equations in real space and describe the dynamics of the free photon,
\begin{equation}\label{eq:vacuum}
\mathcal{H}_0(\mathbf{p})=p_x\hat{S}_x +p_y\hat{S}_y=\begin{bmatrix}
0 & 0 & -p_y\\
0& 0 & p_x\\
-p_y &  p_x & 0 \\
\end{bmatrix}.
\end{equation}
$\mathbf{p}=-i\nabla$ is the two-dimensional momentum operator. $\hat{S}_x$ and $\hat{S}_y$ are spin-1 operators that satisfy the angular momentum algebra $[\hat{S}_i,\hat{S}_j]=i\epsilon_{ijk}\hat{S}_k$,
\begin{equation}\label{eq:Spin}
\hat{S}_z=\begin{bmatrix}
0 & -i & 0 \\
i & 0 & 0 \\
0 & 0 & 0
\end{bmatrix}.
\end{equation}
Here, $(\hat{S}_z)_{ij}=-i\epsilon_{ijz}$ is the generator of rotations in the $x$-$y$ plane and is represented by the antisymmetric matrix. In two dimensions, $\hat{S}_z$ governs all rotational symmetries of the electromagnetic field.

\subsection{2+1D linear response theory}

The effective electromagnetic properties of a material are very accurately described by a linear response theory - assuming nonlinear interactions are negligible. This is true for low intensity light $|f|\lessapprox 10^8~ \mathrm{V/m}$ that is sufficiently weak compared to the atomic fields governing the binding of the crystal itself. Our goal is to characterize the entire topological field theory in this regime. With this in mind, the most general linear response of a 2D material is nonlocal in both space and time coordinates,
\begin{equation}
g(t,\mathbf{r})=  \int d^2\mathbf{r}' \int_{-\infty}^t dt'\mathcal{M}(t,t',\mathbf{r},\mathbf{r}') f(t',\mathbf{r}').
\end{equation}
$\mathcal{M}$ is the response function and compactly represents the constitutive relations in space-time,
\begin{equation}
\mathcal{M}(t,t',\mathbf{r},\mathbf{r}')=\begin{bmatrix}
\varepsilon_{xx} & \varepsilon_{xy} & \chi_{x} \\
\varepsilon_{yx} & \varepsilon_{yy} & \chi_{y}\\
\zeta_{x}& \zeta_{y} & \mu
\end{bmatrix}.
\end{equation}
Note that $\mathcal{M}$ is a $3\times 3$ dimensional matrix and we include all possible material responses as a generalization, for instance magnetism $\mu$ and magnetoelectricity $\chi_i,~\zeta_i$.

If the properties of the crystal are not changing temporally (no external modulation), the response function is translationally invariant in time,
\begin{equation}\label{eq:time_translation}
\begin{split}
\mathcal{M}(t,t',\mathbf{r},\mathbf{r}')&=\mathcal{M}(t-t',\mathbf{r},\mathbf{r}')\\
&=\int d\omega \mathcal{M}(\omega,\mathbf{r},\mathbf{r}')e^{-i\omega(t-t')}.
\end{split}
\end{equation}
Equation~(\ref{eq:time_translation}) implies energy conservation in Hermitian systems $\omega'=\omega$. However, a crystal is not translationally invariant in space - momentum is not conserved $\mathbf{k}'\neq \mathbf{k}$. Instead, the crystal is periodic and possesses \textit{discrete} translational symmetry \cite{agranovich2013crystal,born1954dynamical},
\begin{equation}
\mathcal{M}(\omega,\mathbf{r},\mathbf{r}')=\mathcal{M}(\omega,\mathbf{r}+\mathbf{R},\mathbf{r}'+\mathbf{R}),
\end{equation}
where $\mathbf{R}$ is the primitive lattice vector of the crystal. This admits a Fourier decomposition in the spatial harmonics of the crystal $\mathbf{g}$,
\begin{equation}
\mathcal{M}(\omega,\mathbf{r},\mathbf{r}')=\sum_\mathbf{g}\mathcal{M}_\mathbf{g}(\omega,\mathbf{r}-\mathbf{r}')e^{-i\mathbf{r}'\cdot\mathbf{g}},
\end{equation}
with $\mathbf{g}\cdot\mathbf{R} \in 2\pi~ \mathbb{Z}$ arbitrary integer combinations of the reciprocal lattice vectors.

Due to nonlocality, it is necessary to convert to the reciprocal space,
\begin{equation}
\mathcal{M}(\omega,\mathbf{k},\mathbf{k}')=\iint d^2\mathbf{r} d^2\mathbf{r}'\mathcal{M}(\omega,\mathbf{r},\mathbf{r}')e^{-i\mathbf{k}\cdot\mathbf{r}}e^{i\mathbf{k}'\cdot\mathbf{r}'}.
\end{equation}
$\mathcal{M}(\omega,\mathbf{k},\mathbf{k}')$ determines the linear transformation properties of an input wave with momentum $\mathbf{k}'$ to an output wave with momentum $\mathbf{k}$. In a periodic crystal, the momentum is conserved up to a reciprocal vector $\mathbf{k}'=\mathbf{k}+\mathbf{g}$ and represents a discrete spectrum,
\begin{equation}
\mathcal{M}(\omega,\mathbf{k},\mathbf{k}')=\sum_\mathbf{g}\mathcal{M}_\mathbf{g}(\omega,\mathbf{k})\delta^2(\mathbf{k}+\mathbf{g}-\mathbf{k}').
\end{equation}
$\delta^2(\mathbf{k}+\mathbf{g}-\mathbf{k}')$ is the momentum conserving delta function. Each Fourier element of the response function $\mathcal{M}_\mathbf{g}(\omega,\mathbf{k})$ determines the polarization dependent scattering amplitude from $\mathbf{k}+\mathbf{g}\to\mathbf{k}$. These are essentially the photonic structure factors of the two-dimensional crystal.

In this case, $\mathbf{k}$ is the crystal momentum and is only uniquely defined within the Brillouin zone (BZ). Hence, the electromagnetic eigenstates of the medium are Bloch waves,
\begin{equation}\label{eq:Bloch}
\begin{split}
\mathcal{H}_0(\mathbf{k})f_\mathbf{k}&=\omega\int d^2\mathbf{k}'\mathcal{M}(\omega,\mathbf{k},\mathbf{k}')f_{\mathbf{k}'}\\
&=\omega\sum_\mathbf{g}\mathcal{M}_\mathbf{g}(\omega,\mathbf{k})f_{\mathbf{k}+\mathbf{g}}.
\end{split}
\end{equation}
$\mathcal{H}_0(\mathbf{k})=\mathbf{k}\cdot\mathbf{S}$ are the vacuum Maxwell equations in momentum space. The Bloch photonic wave function $f(\mathbf{k},\mathbf{r})=\langle \mathbf{r}|f_\mathbf{k}\rangle$ corresponds to the net propagation of all $\mathbf{k}+\mathbf{g}$ scattered waves in the medium, 
\begin{equation}\label{eq:BlochPhoton}
f(\mathbf{k},\mathbf{r})=\sum_\mathbf{g} f_{\mathbf{k}+\mathbf{g}}e^{i\mathbf{g}\cdot\mathbf{r}},
\end{equation}
where $f(\mathbf{k},\mathbf{r}+\mathbf{R})=f(\mathbf{k},\mathbf{r})$ is periodic in the crystal lattice. Note that Eq.~(\ref{eq:Bloch}) and (\ref{eq:BlochPhoton}) reduce to the continuum theory \cite{VanMechelen2018,van_mechelen_photonic_2018} when considering only the 0th order harmonic $\mathbf{g}=0$.

\subsection{Generalized response function}

Nevertheless, Eq.~(\ref{eq:Bloch}) poses a few serious problems; it does not represent a proper first-order in time Hamiltonian since all harmonics of the response function $\mathcal{M}_\mathbf{g}(\omega,\mathbf{k})$ depend on the eigenvalue $\omega$. Moreover, it is not evident that the Bloch waves in Eq.~(\ref{eq:BlochPhoton}) are normalizable, as the system contains complex spatial and temporal dispersion. Due to these issues, it is advantageous to return to the more general form of $\mathcal{M}(\omega,\mathbf{k},\mathbf{k}')$ without assuming discrete translational symmetry. This will allow us to derive very robust properties of the response function that can also be applied to amorphous materials or quasicrystals.

First, we demand Hermiticity,
\begin{equation}\label{eq:Hermiticity}
\mathcal{M}(\omega,\mathbf{k},\mathbf{k}')=\mathcal{M}^\dagger(\omega,\mathbf{k}',\mathbf{k}),
\end{equation}
such that the response is lossless. To account for normalizable electromagnetic waves, the energy density must be positive definite for all $\omega$,
\begin{equation}\label{eq:Energy_density}
U(\omega)=\iint d^2\mathbf{k}d^2\mathbf{k}'f^\dagger_\mathbf{k}\bar{\mathcal{M}}(\omega,\mathbf{k},\mathbf{k}')f_{\mathbf{k}'}>0,
\end{equation}
where $\bar{\mathcal{M}}$ describes the inner product space in a dispersive medium,
\begin{equation}
\bar{\mathcal{M}}(\omega,\mathbf{k},\mathbf{k}')=\frac{\partial}{\partial\omega}\left[\omega\mathcal{M}(\omega,\mathbf{k},\mathbf{k}')\right].
\end{equation}
Notice that $U(\omega)=U^*(\omega)$ is only real-valued when $\mathcal{M}$ is Hermitian. For realistic materials, the energy density is also stable at static equilibrium $\omega=0$,
\begin{equation}
U(0)=\iint d^2\mathbf{k}d^2\mathbf{k}'f^\dagger_\mathbf{k}\mathcal{M}(0,\mathbf{k},\mathbf{k}')f_{\mathbf{k}'}>0,
\end{equation}
with $\mathcal{M}(0,\mathbf{k},\mathbf{k}')=\bar{\mathcal{M}}(0,\mathbf{k},\mathbf{k}')$ at zero frequency. To ensure the electromagnetic field is real-valued, i.e. represents a neutral particle, we always require the reality condition,
\begin{equation}
\mathcal{M}(\omega,\mathbf{k},\mathbf{k}')=\mathcal{M}^*(-\omega,-\mathbf{k},-\mathbf{k}').
\end{equation}
Furthermore, the response is transparent at high frequency $\omega\to\infty$, as the material cannot respond to sufficiently fast temporal oscillations,
\begin{equation}
\lim_{\omega\to\infty}\mathcal{M}(\omega,\mathbf{k},\mathbf{k}')=\mathds{1}_3\delta^2_{\mathbf{k}-\mathbf{k}'}.
\end{equation}
$\mathds{1}_3$ is the $3\times 3$ identity matrix and $\delta^2_{\mathbf{k}-\mathbf{k}'}=\delta^2(\mathbf{k}-\mathbf{k}')$ is the momentum conserving delta function. Lastly, the response must be causal and satisfy the Kramers-Kronig relations.

Combining all the above criteria, we find that $\mathcal{M}$ can always be decomposed as a discrete summation of oscillators \cite{Haldane2008,Silveirinha2015,Philbin2014},
\begin{equation}\label{eq:Response}
\mathcal{M}(\omega,\mathbf{k},\mathbf{k}')=\mathds{1}_3\delta^2_{\mathbf{k}-\mathbf{k}'}-\sum_\alpha\int d^2\mathbf{k}''\frac{\mathcal{C}^\dagger_{\alpha\mathbf{k}''\mathbf{k}}\mathcal{C}_{\alpha\mathbf{k}''\mathbf{k}'}}{\omega_{\alpha\mathbf{k}''}(\omega-\omega_{\alpha\mathbf{k}''})}.
\end{equation}
Any Hermitian (lossless) response function can be expressed in this form. Equation~(\ref{eq:Response}) is easily extended to 3D materials but our focus is on 2D topological field theories. In this case, $\alpha$ labels an arbitrary bosonic excitation in the material, such as an exciton or phonon, which couples linearly to the electromagnetic fields via the $3\times 3$ tensor,
\begin{equation}
\mathcal{C}_\alpha(\mathbf{k},\mathbf{k}')=\iint d^2\mathbf{r} d^2\mathbf{r}'\mathcal{C}_\alpha(\mathbf{r},\mathbf{r}')e^{-i\mathbf{k}\cdot\mathbf{r}}e^{i\mathbf{k}'\cdot\mathbf{r}'}.
\end{equation}
$\omega_{\alpha\mathbf{k}}$ is the resonant energy of the oscillator and corresponds to a first-order pole of the response function. Notice that $\mathcal{M}$ itself contains an integral over $\mathbf{k}''$. Microscopically, this constitutes the overlap with the electronic momentum to infinitesimally small scale $k\to \infty$.

Substituting Eq.~(\ref{eq:Response}) into Eq.~(\ref{eq:Energy_density}), we can exchange the order of integration $U(\omega)=\int d^2\mathbf{k}U(\omega,\mathbf{k})$ and define,
\begin{equation}\label{eq:Energy_density2}
U(\omega,\mathbf{k})=|f_\mathbf{k}|^2+\sum_\alpha \left|\int d^2\mathbf{k}'\frac{\mathcal{C}_{\alpha\mathbf{k}\mathbf{k}'} f_{\mathbf{k}'}}{(\omega-\omega_{\alpha\mathbf{k}})}\right|^2>0,
\end{equation}
which is positive definite for all $\omega$ and $\mathbf{k}$. Equation~(\ref{eq:Energy_density2}) is the generalized inner product for the electromagnetic field and represents the energy density at an arbitrary frequency and wave vector. We will now show that Eq.~(\ref{eq:Response}) is derived from a first-order in time Hamiltonian.

\subsection{Generalized Hamiltonian}

To find the corresponding Hamiltonian, we expand the response function $\mathcal{M}$ in terms of three-component matter oscillators $\psi_\alpha$. These represent internal polarization and magnetization modes of the material, 
\begin{equation}\label{eq:oscillator}
\omega \psi_{\alpha\mathbf{k}}=\omega_{\alpha\mathbf{k}}\psi_{\alpha\mathbf{k}}+\int d^2\mathbf{k}'\mathcal{C}_{\alpha\mathbf{k}\mathbf{k}'}f_{\mathbf{k}'}.
\end{equation}
Substituting Eq.~(\ref{eq:oscillator}) and (\ref{eq:Response}) into Eq.~(\ref{eq:Bloch}) we obtain,
\begin{equation}\label{eq:self}
\begin{split}
\omega f_\mathbf{k}=\mathcal{H}_0(\mathbf{k}) f_\mathbf{k}+&\sum_\alpha\iint  \frac{d^2\mathbf{k}''d^2\mathbf{k}'}{\omega_{\alpha\mathbf{k}''}}\mathcal{C}^\dagger_{\alpha\mathbf{k}''\mathbf{k}}\mathcal{C}_{\alpha\mathbf{k}''\mathbf{k}'} f_{\mathbf{k}'}\\+&\sum_\alpha \int d^2\mathbf{k}' \mathcal{C}^\dagger_{\alpha\mathbf{k}'\mathbf{k}}\psi_{\alpha\mathbf{k}'}.
\end{split}
\end{equation}
The first two terms on the right hand side of Eq.~(\ref{eq:self}) represent the vacuum equations and self-energy of the electromagnetic field. The third term is the linear coupling to the oscillators. Combining Eq.~(\ref{eq:oscillator}) and (\ref{eq:self}) into a single algebraic matrix, we write the generalized Hamiltonian $H(\mathbf{k},\mathbf{k}')$ as,
\begin{widetext}
\begin{equation}\label{eq:Hamiltonian}
H(\mathbf{k},\mathbf{k}')=\begin{bmatrix}
\mathcal{H}_0(\mathbf{k})\delta^2_{\mathbf{k}-\mathbf{k}'}+\sum_\alpha\int  \frac{d^2\mathbf{k}''}{\omega_{\alpha\mathbf{k}''}}\mathcal{C}^\dagger_{\alpha\mathbf{k}''\mathbf{k}}\mathcal{C}_{\alpha\mathbf{k}''\mathbf{k}'} & ~~\mathcal{C}^\dagger_{1\mathbf{k}'\mathbf{k}}~~ & ~~\mathcal{C}^\dagger_{2\mathbf{k}'\mathbf{k}}~~ & \ldots \\
\mathcal{C}_{1\mathbf{k}\mathbf{k}'} & \omega_{1\mathbf{k}}\delta^2_{\mathbf{k}-\mathbf{k}'} & 0 & \ldots\\
\mathcal{C}_{2\mathbf{k}\mathbf{k}'} & 0 & \omega_{2\mathbf{k}}\delta^2_{\mathbf{k}-\mathbf{k}'} & \ldots\\
\vdots & \vdots & \vdots & \ddots 
\end{bmatrix},
\end{equation}
\end{widetext}
which is manifestly Hermitian $H(\mathbf{k},\mathbf{k}')=H^\dagger(\mathbf{k}',\mathbf{k})$.

We now define $u_\mathbf{k}$ as the generalized state vector of the electromagnetic problem; accounting for the photon $f_\mathbf{k}$ and all possible internal excitations $\psi_{\alpha\mathbf{k}}$,
\begin{equation}
\int d^2\mathbf{k}' H_{\mathbf{k}\mathbf{k}'}u_{\mathbf{k}'}=\omega u_\mathbf{k}, \qquad
u_\mathbf{k}=\begin{bmatrix}
f_\mathbf{k} \\ \psi_{1\mathbf{k}} \\ \psi_{2\mathbf{k}} \\ \vdots
\end{bmatrix},
\end{equation}
which is a first-order wave equation. Notice that contraction of $u_\mathbf{k}$ naturally reproduces the energy density [Eq.~(\ref{eq:Energy_density2})] upon summation over all degrees of freedom,
\begin{equation}\label{eq:Normalization}
\begin{split}
u^\dagger_\mathbf{k} u_\mathbf{k}&=|f_\mathbf{k}|^2+\sum_\alpha |\psi_{\alpha\mathbf{k}}|^2=U(\omega,\mathbf{k})\\
&=|f_\mathbf{k}|^2+\sum_\alpha \left|\int d^2\mathbf{k}'\frac{\mathcal{C}_{\alpha\mathbf{k}\mathbf{k}'} f_{\mathbf{k}'}}{(\omega-\omega_{\alpha\mathbf{k}})}\right|^2.
\end{split}
\end{equation}
The complete set of eigenvectors and eigenvalues is represented by $u_\mathbf{k}$. We must define all relevant electromagnetic quantities in terms of this generalized state vector. 

\subsection{Crystal Hamiltonian}

We are now ready to enforce crystal periodicity. Instead of expanding $\mathcal{M}$ directly, we utilize the periodicity of the coupling tensors $\mathcal{C}_\alpha(\mathbf{r},\mathbf{r}')=\mathcal{C}_\alpha(\mathbf{r}+\mathbf{R},\mathbf{r}'+\mathbf{R})$, which is a discrete spectrum in $\mathbf{g}$,
\begin{equation}
\mathcal{C}_\alpha(\mathbf{k},\mathbf{k}')=\sum_\mathbf{g}\mathcal{C}_{\alpha\mathbf{g}}(\mathbf{k})\delta^2(\mathbf{k}+\mathbf{g}-\mathbf{k}').
\end{equation}
$\mathcal{C}_{\alpha\mathbf{g}}(\mathbf{k})$ tells us the scattering amplitude of a photon $f_{\mathbf{k}+\mathbf{g}}$ with momentum $\mathbf{k}+\mathbf{g}$ into an internal mode of the material $\psi_{\alpha\mathbf{k}}$ at momentum $\mathbf{k}$, and vice versa. The crystal Hamiltonian accounts for all such scattering events, 
\begin{equation}\label{eq:FourierHamiltonian}
H(\mathbf{k},\mathbf{k}')=\sum_\mathbf{g}H_{\mathbf{g}}(\mathbf{k})\delta^2(\mathbf{k}+\mathbf{g}-\mathbf{k}'),
\end{equation}
with Hermiticity $H_\mathbf{g}(\mathbf{k})=H^\dagger_{-\mathbf{g}}(\mathbf{k}+\mathbf{g})$ satisfied by definition. Note, the resonant energies $\omega_{\alpha}(\mathbf{k}+\mathbf{g})=\omega_\alpha(\mathbf{k})$ are generally periodic in $\mathbf{k}$, since they correspond to energy gaps in the electronic band structure. However, we do not need to assume this to define the optical Bloch excitations. A periodic coupling is sufficient.

The quasiparticle eigenstates of this Hamiltonian describe the complete spectrum of Bloch waves,
\begin{equation}\label{eq:DiscreteHamiltonian}
\sum_\mathbf{g}H_\mathbf{g}(\mathbf{k})u_{n\mathbf{k}+\mathbf{g}}=\omega_{n\mathbf{k}}u_{n\mathbf{k}}, \qquad \omega_{n}(\mathbf{k}+\mathbf{g})=\omega_{n}(\mathbf{k}),
\end{equation}
and the eigenenergies $\omega_{n\mathbf{k}}$ are periodic Bloch bands. $n$ labels a particular energy band of the material with its associated Bloch eigenstate $|u_{n\mathbf{k}}\rangle$. The total wave function $|u_{n\mathbf{k}}\rangle$ contains the photon $|f_{n\mathbf{k}}\rangle$ and all internal degrees of freedom describing the linear response $|\psi_{n\alpha\mathbf{k}}\rangle$. This is expressed compactly in the Fourier basis $u_n(\mathbf{k},\mathbf{r})=\langle \mathbf{r}|u_{n\mathbf{k}}\rangle$,
\begin{equation}
u_n(\mathbf{k},\mathbf{r})=\sum_\mathbf{g}u_{n\mathbf{k}+\mathbf{g}}e^{i\mathbf{g}\cdot\mathbf{r}}, \qquad u_{n\mathbf{k}+\mathbf{g}}=\begin{bmatrix}
f_{n\mathbf{k}+\mathbf{g}} \\ \psi_{n1\mathbf{k}+\mathbf{g}} \\ \psi_{n2\mathbf{k}+\mathbf{g}} \\ \vdots
\end{bmatrix},
\end{equation}
where $u_n(\mathbf{k},\mathbf{r}+\mathbf{R})=u_n(\mathbf{k},\mathbf{r})$ is periodic in the crystal lattice. In this basis, $|u_{n\mathbf{k}}\rangle$ is normalized to the energy density as,
\begin{equation}\label{eq:BlochNormalization}
\begin{split}
1&=\langle u_{n\mathbf{k}}|u_{n\mathbf{k}}\rangle =\sum_\mathbf{g}u^\dagger_{n\mathbf{k}+\mathbf{g}}u_{n\mathbf{k}+\mathbf{g}}\\
&=\sum_\mathbf{g}\left(f^\dagger_{n\mathbf{k}+\mathbf{g}}f_{n\mathbf{k}+\mathbf{g}}+\sum_\alpha\psi^\dagger_{n\alpha\mathbf{k}+\mathbf{g}}\psi_{n\alpha\mathbf{k}+\mathbf{g}}\right)\\
&=\sum_\mathbf{g\mathbf{g}'}f^\dagger_{n\mathbf{k}+\mathbf{g}}\bar{\mathcal{M}}_{\mathbf{g}'-\mathbf{g}}(\omega_{n\mathbf{k}},\mathbf{k}+\mathbf{g})f_{n\mathbf{k}+\mathbf{g}'}.
\end{split}
\end{equation}
The bra-ket notation $\langle|\rangle$ implies integration over the 2D unit cell and we have utilized the linear response theory to express $\psi_{\alpha}$ in terms of the driving field $f$,
\begin{equation}
\psi_{n\alpha\mathbf{k}+\mathbf{g}}=\frac{\sum_{\mathbf{g}'}\mathcal{C}_{\alpha\mathbf{g}'}(\mathbf{k}+\mathbf{g})f_{n\mathbf{k}+\mathbf{g}'+\mathbf{g}}}{\omega_{n\mathbf{k}}-\omega_{\alpha\mathbf{k}+\mathbf{g}}}.
\end{equation}
$\bar{\mathcal{M}}_\mathbf{g}(\omega,\mathbf{k})=\partial_\omega\left[\omega\mathcal{M}_\mathbf{g}(\omega,\mathbf{k})\right]$ is the contribution to the energy density arising from each spatial harmonic of the crystal.

Finally, the eigenenergies $\omega_{n\mathbf{k}}$ are the $n$ nontrivial roots of the characteristic wave equation,
\begin{equation}
\mathcal{H}_0(\mathbf{k})f_{n\mathbf{k}}=\omega_{n\mathbf{k}}\sum_\mathbf{g}\mathcal{M}_\mathbf{g}(\omega_{n\mathbf{k}},\mathbf{k})f_{n\mathbf{k}+\mathbf{g}},
\end{equation}
which generates all possible photonic bands of the crystal. Note, the response function $\mathcal{M}_{\mathbf{g}}(\omega,\mathbf{k})$ is now expressed in terms of $\mathcal{C}_{\alpha\mathbf{g}}(\mathbf{k})$ and describes the net summation of all scattering and back-scattering events in the material,
\begin{equation}\label{eq:ResponseFourier}
\mathcal{M}_{\mathbf{g}}(\omega,\mathbf{k})=\mathds{1}_3\delta_{\mathbf{g}}-\sum_{\alpha\mathbf{g}'}\frac{\mathcal{C}^\dagger_{\alpha-\mathbf{g}'}(\mathbf{k}+\mathbf{g}')\mathcal{C}_{\alpha\mathbf{g}-\mathbf{g}'}(\mathbf{k}+\mathbf{g}')}{\omega_{\alpha\mathbf{k}+\mathbf{g}'}(\omega-\omega_{\alpha\mathbf{k}+\mathbf{g}'})}.
\end{equation}
This proves that the wave equation is derived from a first-order Hamiltonian, has real eigenvalues $\omega=\omega_{n\mathbf{k}}$ for all momenta, and is normalizable in terms of $|u_{n\mathbf{k}}\rangle$.

\section{Discrete rotational symmetry}\label{sec:Rotational_Symmetry}

\subsection{Point groups in 2D}

\begin{figure}
  \includegraphics[width=0.9\columnwidth]{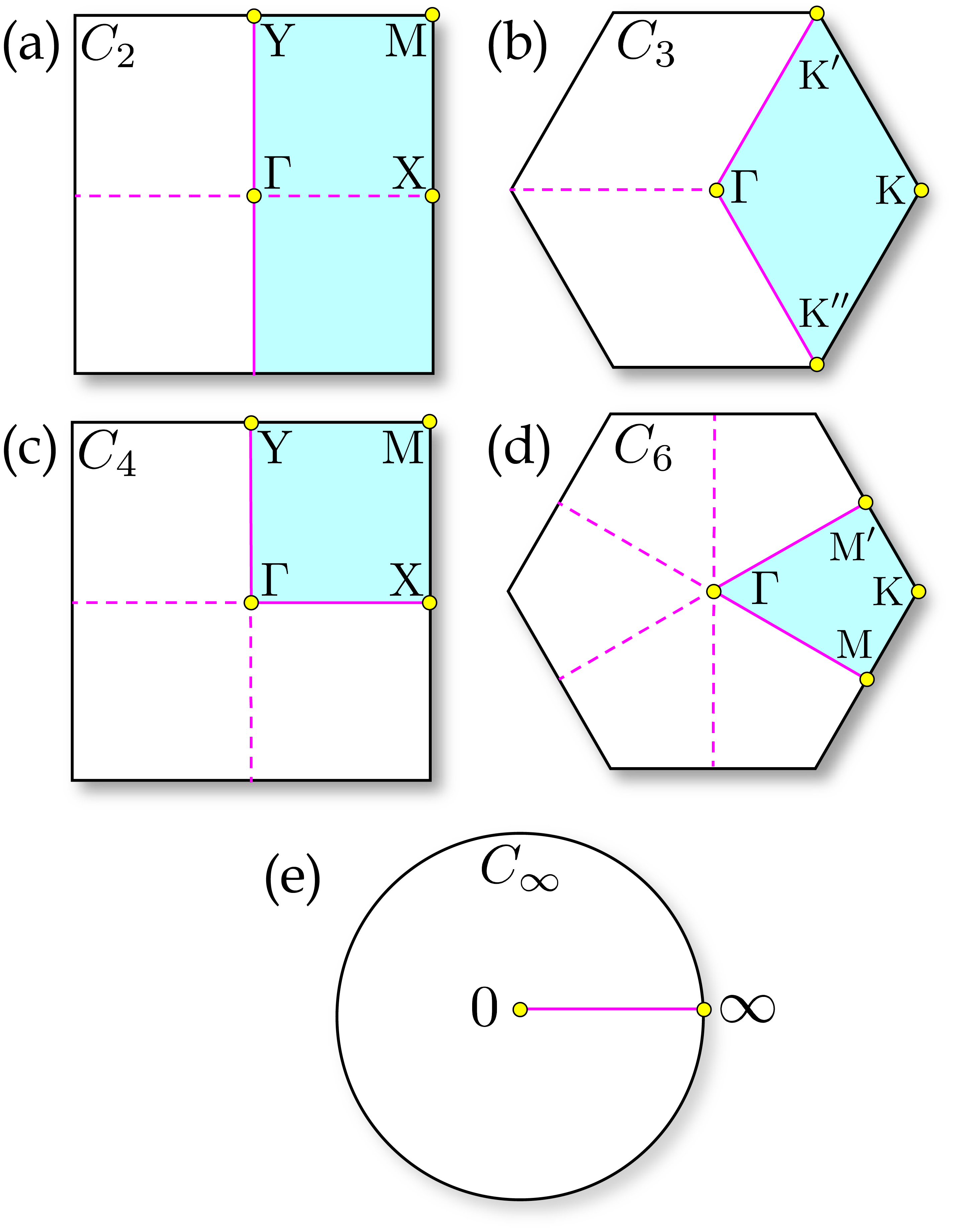}
  \caption{Brillouin zone of each cyclic point group $C_N$. (a), (b), (c), (d), and (e) correspond to $N=2,3,4,6$, and $\infty$ respectively. Due to rotational symmetry, the total Brillouin zone is equivalent to $N$ copies of the irreducible Brillouin zone ($\textrm{IBZ}$), which is represented by the blue quadrant. For continuous symmetry $N=\infty$, this is simply a line. The yellow circles label high-symmetry points $\mathcal{R}\mathbf{k}_i=\mathbf{k}_i$ where the crystal Hamiltonian is invariant under a certain rotation $\hat{\mathcal{R}}$. At these specific momenta, a Bloch photonic wave function $\hat{\mathcal{R}}_N|f(\mathbf{k}_i)\rangle=\eta_{N}(\mathbf{k}_i)|f(\mathbf{k}_i)\rangle$ is an eigenstate of a an $N$-fold rotation $\eta_{N}(\mathbf{k}_i)=\left[i\frac{2\pi}{N} m_N(\mathbf{k}_i)\right]$ such that the photon possesses quantized integer eigenvalues $m_N(\mathbf{k}_i)\in \mathbb{Z}_N$. Since $m_N$ are discrete quantum numbers, their values cannot vary continuously if the crystal symmetry is preserved - they can only be changed at a topological phase transition.}
    \label{fig:PointGroups}
\end{figure}

Point groups are the discrete analogs of continuous rotations and reflections. They represent the number of ways the atomic lattice can be transformed into itself \cite{Hestenes2002,nye1985physical}. Due to the crystallographic restriction theorem (CRT), there are ten such point groups in 2D. The first five are the cyclic groups $C_N$,
\begin{equation}
C_1,~C_2,~C_3,~C_4,~C_6.
\end{equation}
For instance, $C_3$ implies threefold cyclic symmetry while $C_1$ is no symmetry. The last five are the dihedral groups $D_N$,
\begin{equation}
D_1,~D_2,~D_3,~D_4,~D_6.
\end{equation}
The dihedral group $D_N$ contains $C_N$ plus reflections. However, it can be proven that the Chern number for all $D_N$ point groups vanish \cite{Fang2012}. Therefore, we concern ourselves with only the cyclic groups $C_N$. The Brillouin zone of each point group is displayed in Fig.~\ref{fig:PointGroups}.

The defining characteristic of each cyclic group is the fermionic or bosonic representation. When we rotate the fields by $2\pi$, we take the particle into itself and acquire a phase,
\begin{equation}
\mathcal{R}(2\pi)=(-1)^F.
\end{equation}
$F$ is twice the total spin of particle, or equivalently, the fermion number. Fermions with half-integer spin are antisymmetric under rotations $\mathcal{R}(2\pi)=-1$, while bosons with integer spin are symmetric $\mathcal{R}(2\pi)=+1$. Depending on the symmetries of the lattice, the topology fundamentally changes for fermions and bosons. We will understand the implications this has for spin-1 photons.

\subsection{Spin-1 discrete symmetries}

If the two-dimensional crystal belongs to a cyclic point group $C_N$, the Hamiltonian possesses discrete rotational symmetry about the $z$-axis,
\begin{equation}\label{eq:rotational}
\mathcal{R}^{-1}H_{\mathcal{R}\mathbf{g}}(\mathcal{R}\mathbf{k})\mathcal{R}=H_\mathbf{g}(\mathbf{k}), \qquad \omega_n(\mathcal{R}\mathbf{k})=\omega_n(\mathbf{k}),
\end{equation}
where $\mathcal{R}$ is any rotation in $C_N$. It is important to note that $\mathcal{R}$ is \textit{diagonal} in $u$, meaning the photon and each oscillator is rotated individually, $f\to \mathcal{R}f$ and $\psi_\alpha\to \mathcal{R}\psi_\alpha$. This implies there is no mixing of  fields. The symmetries of the Hamiltonian are endowed by the coupling tensors, which dictates the degrees of freedom of the material response,
\begin{equation}\label{eq:rotational2}
\mathcal{R}^{-1}\mathcal{C}_{\alpha\mathcal{R}\mathbf{g}}(\mathcal{R}\mathbf{k})\mathcal{R}=\mathcal{C}_{\alpha\mathbf{g}}(\mathbf{k}), \qquad \omega_\alpha(\mathcal{R}\mathbf{k})=\omega_\alpha(\mathbf{k}).
\end{equation}
After summation over all $\mathcal{C}_{\alpha\mathbf{g}}(\mathbf{k})$, we can prove that the response function transforms identically under such a rotation,
\begin{equation}\label{eq:rotational3}
\mathcal{R}^{-1}\mathcal{M}_{\mathcal{R}\mathbf{g}}(\omega,\mathcal{R}\mathbf{k})\mathcal{R}=\mathcal{M}_\mathbf{g}(\omega,\mathbf{k}).
\end{equation}
Therefore, the photon inherits all symmetries of the crystal.

In this case, the $\mathcal{R}$ matrix represents a discrete rotation and can be expressed as the exponential of the spin-1 generator $(\hat{S}_z)_{ij}=-i\epsilon_{ijz}$,
\begin{equation}\label{eq:RotationMatrix}
\mathcal{R}_N=\exp\left(i\frac{2\pi}{N}\hat{S}_z\right)=\begin{bmatrix}
\cos\frac{2\pi}{N} & \sin\frac{2\pi}{N} & 0 \\
-\sin\frac{2\pi}{N} & \cos\frac{2\pi}{N} & 0 \\
0 & 0 &1
\end{bmatrix},
\end{equation}
where $\sfrac{2\pi}{N}$ is an $N$-fold rotation. We stress that every cyclic group for the photon is a vector representation, which is \textit{bosonic},
\begin{equation}
\mathcal{R}(2\pi)=\mathcal{R}(N\theta_N)=(\mathcal{R}_N)^N=+\mathds{1}_3.
\end{equation}
The electromagnetic field returns in phase under cyclic revolution.

\subsection{High-symmetry points}

The Bloch eigenstates $|u_{n\mathbf{k}}\rangle$ are essentially a collection of periodic vector fields. To rotate the fields, we must perform an operation on both the coordinates $\mathbf{r}$ and the polarization states $f$ and $\psi_\alpha$. In real space, the operation of a rotation $\hat{\mathcal{R}}$ is preformed as,
\begin{equation}
\langle \mathbf{r}|\hat{\mathcal{R}}|u_{n\mathbf{k}}\rangle=\mathcal{R}u_n(\mathbf{k},\mathcal{R}^{-1}\mathbf{r})=\eta_{n}(\mathbf{k}) u_n(\mathcal{R}\mathbf{k},\mathbf{r}),
\end{equation}
where $\mathcal{R}$ is a discrete rotation defined in Eq.~(\ref{eq:RotationMatrix}). This implies the Fourier coefficients obey,
\begin{equation}
\mathcal{R}u_{n\mathbf{k}+\mathcal{R}^{-1}\mathbf{g}}=\eta_n(\mathbf{k})u_{n\mathcal{R}\mathbf{k}+\mathbf{g}}.
\end{equation}
It follows from symmetry that the operation of $\hat{\mathcal{R}}$ takes a wave function at $\mathbf{k}$ to $\mathcal{R}\mathbf{k}$ with the same energy $\omega_{n}(\mathbf{k})=\omega_n(\mathcal{R}\mathbf{k})$ - but with a possibly different phase $|\eta_{n}(\mathbf{k})|^2=1$. Utilizing the linear response theory, we notice that the phase factor $\eta_n(\mathbf{k})$ is governed entirely by the photon,
\begin{equation}
\begin{split}
\mathcal{R}\psi_{n\alpha\mathbf{k}+\mathcal{R}^{-1}\mathbf{g}}&=\frac{\sum_{\mathbf{g}'}\mathcal{R}\mathcal{C}_{\alpha\mathbf{g}'}(\mathbf{k}+\mathcal{R}^{-1}\mathbf{g})f_{n\mathbf{k}+\mathbf{g}'+\mathcal{R}^{-1}\mathbf{g}}}{\omega_{n\mathbf{k}}-\omega_{\alpha\mathbf{k}+\mathcal{R}^{-1}\mathbf{g}}}\\
&=\frac{\sum_{\mathbf{g}'}\mathcal{C}_{\alpha\mathcal{R}\mathbf{g}'}(\mathcal{R}\mathbf{k}+\mathbf{g})\mathcal{R}f_{n\mathbf{k}+\mathbf{g}'+\mathcal{R}^{-1}\mathbf{g}}}{\omega_{n\mathbf{k}}-\omega_{\alpha\mathbf{k}+\mathcal{R}^{-1}\mathbf{g}}}\\
&=\frac{\sum_{\mathbf{g}'}\mathcal{C}_{\alpha\mathbf{g}'}(\mathcal{R}\mathbf{k}+\mathbf{g})\eta_n(\mathbf{k})f_{n\mathcal{R}\mathbf{k}+\mathbf{g}'+\mathbf{g}}}{\omega_{n\mathbf{k}}-\omega_{\alpha\mathcal{R}\mathbf{k}+\mathbf{g}}}\\
&=\eta_n(\mathbf{k})\psi_{n\alpha\mathcal{R}\mathbf{k}+\mathbf{g}}.
\end{split}
\end{equation}
This is an incredibly convenient simplification and implies the precise coordinates of the matter oscillations $\psi_\alpha$ are superfluous when discussing symmetries. The electromagnetic field $f$ tells us everything.

Importantly, there are specific points in the Brillouin zone where $\mathbf{k}$ is invariant under a discrete rotation,
\begin{equation}
\mathcal{R}\mathbf{k}_i=\mathbf{k}_i.
\end{equation}
This is because the crystal momentum only differs by a lattice translation at these points $\mathcal{R}\mathbf{k}_i=\mathbf{k}_i+\mathbf{g}$, which leaves a Bloch wave function unchanged, 
\begin{equation}
\begin{split}
e^{i\mathcal{R}\mathbf{k}_i\cdot\mathbf{r}}u_{n}(\mathcal{R}\mathbf{k}_i,\mathbf{r})&=e^{i(\mathbf{k}_i+\mathbf{g})\cdot\mathbf{r}}u_{n}(\mathbf{k}_i+\mathbf{g},\mathbf{r})\\
&=e^{i\mathbf{k}_i\cdot\mathbf{r}}u_{n}(\mathbf{k}_i,\mathbf{r}).
\end{split}
\end{equation}
These are called \textit{high-symmetry points} (HSPs); they occur at the center and certain vertices of the Brillouin zone. The crystal Hamiltonian is \textit{rotationally invariant} at these momenta - i.e. it commutes with $\hat{\mathcal{R}}$. Therefore, the wave functions are simultaneous eigenstates of $\hat{\mathcal{R}}$ at HSPs,
\begin{equation}
\hat{\mathcal{R}}|u_{n}(\mathbf{k}_i)\rangle=\eta_n(\mathbf{k}_i)|u_{n}(\mathbf{k}_i)\rangle,
\end{equation}
which immediately implies,
\begin{equation}
\hat{\mathcal{R}}|f_{n}(\mathbf{k}_i)\rangle=\eta_n(\mathbf{k}_i)|f_{n}(\mathbf{k}_i)\rangle.
\end{equation}
Here, $\eta_n(\mathbf{k}_i)$ is the eigenvalue of $\hat{\mathcal{R}}$ at $\mathbf{k}_i$ for the $n$th band.

\begin{figure*}
  \includegraphics[width=0.9\textwidth]{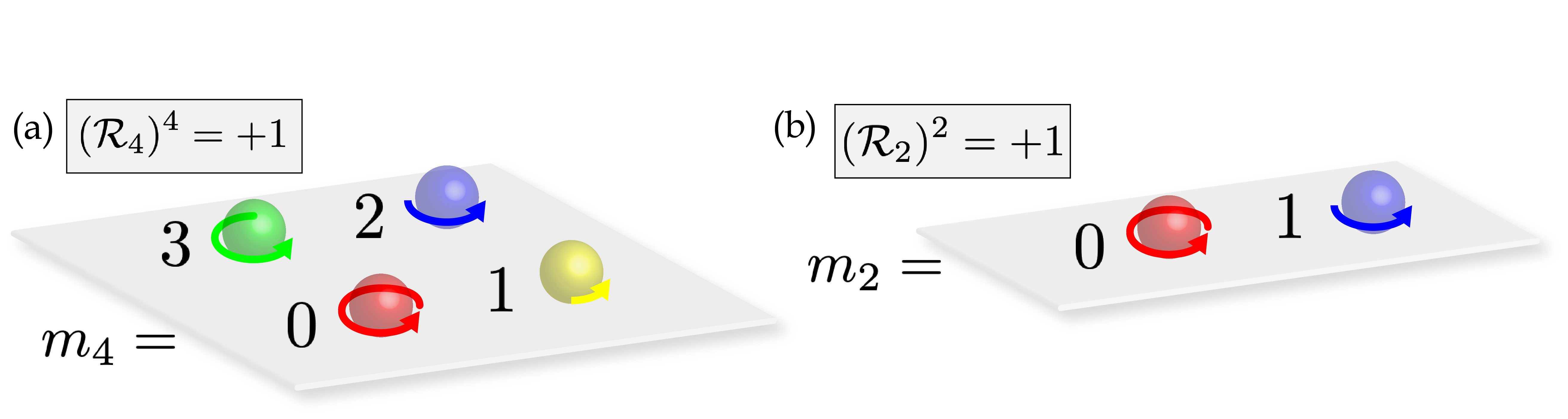}
  \caption{The collection of spin-1 (bosonic) charges for the $C_4$ point group. (a) Fourfold rotations $(\mathcal{R}_4)^4=+1$; there are four unique eigenvalues $\eta_4=\exp\left[i\frac{2\pi}{4}m_4\right]$ corresponding to the roots of unity $(\eta_4)^4=+1$. These represent the modulo 4 integers $m_4\in\mathbb{Z}_4$. Note that $m_4=3=-1$ can also be interpreted as a left-handed eigenstate. (b) Bosonic inversion $(\mathcal{R}_2)^2=+1$; there are two unique eigenvalues $\eta_2=\exp\left[i\frac{2\pi}{2}m_2\right]$ corresponding to the roots of unity $(\eta_2)^2=+1$. These represent the modulo 2 integers $m_2\in\mathbb{Z}_2$.} 
  \label{fig:quantized_states}
\includegraphics[width=0.9\textwidth]{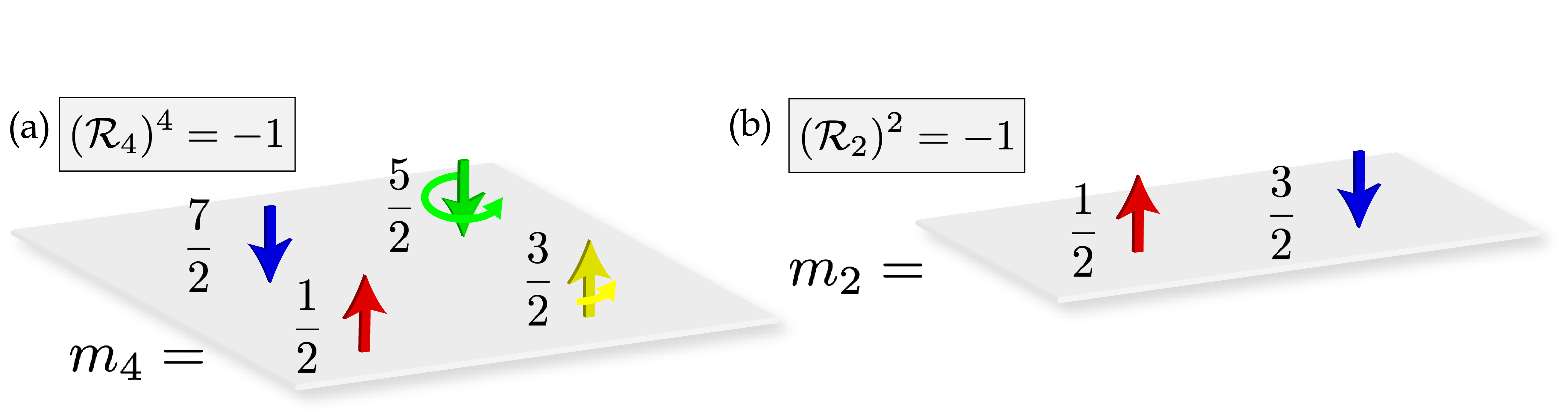}
  \caption{The collection of spin-\sfrac{1}{2} (fermionic) charges for the $C_4$ point group. (a) Fourfold rotations $(\mathcal{R}_4)^4=-1$; there are four unique eigenvalues $\zeta_4=\exp\left[i\frac{2\pi}{4}m_4\right]$ corresponding to the roots of negative unity $(\zeta_4)^4=-1$. These represent the modulo 4 half-integers $m_4\in\mathbb{Z}_4+\sfrac{1}{2}$. Note that $m_4=\sfrac{7}{2}=-\sfrac{1}{2}$ can be interpreted as a spin-down fermion while $m_4=\sfrac{3}{2}=\sfrac{1}{2}+1$ and $m_4=\sfrac{5}{2}=-\sfrac{1}{2}+3$ constitute a fermion plus a boson. (b) Fermionic inversion $(\mathcal{R}_2)^2=-1$; there are two unique eigenvalues $\zeta_2=\exp\left[i\frac{2\pi}{2}m_2\right]$ corresponding to the roots of negative unity $(\zeta_2)^2=-1$. These represent the modulo 2 half-integers $m_2\in\mathbb{Z}_2+\sfrac{1}{2}$. Note that $m_2=\sfrac{3}{2}=-\sfrac{1}{2}$ can also be interpreted as a spin-down fermion under modulo 2.}
  \label{fig:quantized_states_2}
\end{figure*}

\subsection{Spin-1 eigenvalues}

Depending on the point group and the precise HSP, $\eta_{n}(\mathbf{k}_i)=\eta_{N,n}(\mathbf{k}_i)$ can represent any $N$th root of unity corresponding to the rotation operator $\hat{\mathcal{R}}_N$,
\begin{equation}
\eta_{N,n}(\mathbf{k}_i)=\exp\left[i\frac{2\pi}{N} m_{N,n}(\mathbf{k}_i)\right], \qquad (\eta_{N,n})^N=+1.
\end{equation}
$m_{N,n}(\mathbf{k}_p)\in \mathbb{Z}_N$ is a modulo integer - it labels the $N$ possible spin-1 eigenvalues at $\mathbf{k}_i$. In $C_4$ for example, the $\Gamma$ and $\textrm{M}$ points are invariant under $\hat{\mathcal{R}}_4$ rotations, while the $\textrm{X}$ and $\textrm{Y}$ points are invariant under $\hat{\mathcal{R}}_2$ rotations (inversion). This means there are 4 possible spin-1 charges located at $m_{4,n}(\Gamma)$ \& $m_{4,n}(\textrm{M})\in \mathbb{Z}_4$ respectively and 2 possible charges located at $m_{2,n}(\textrm{X})=m_{2,n}(\textrm{Y})\in \mathbb{Z}_2$. A visualization of these topological charges is presented in Fig.~\ref{fig:quantized_states} and this is contrasted with their fermionic counterparts in Fig.~\ref{fig:quantized_states_2}. In Sec. \ref{sec:Bosonic_Phases} we will connect these rotational eigenvalues directly to the topological invariants.

\section{Topological electromagnetic (bosonic) phases of matter}\label{sec:Bosonic_Phases}

\subsection{Electromagnetic Chern number}

The Berry connection for a band $n$ is found by varying the total Bloch wave function $|u_{n\mathbf{k}}\rangle$ with respect to the momentum,
\begin{equation}
\mathbf{A}_{n}(\mathbf{k})=-i\langle u_{n\mathbf{k}}|\partial_\mathbf{k} u_{n\mathbf{k}}\rangle=-i\sum_\mathbf{g}u^\dagger_{n\mathbf{k}+\mathbf{g}}\partial_\mathbf{k} u_{n\mathbf{k}+\mathbf{g}}.
\end{equation}
This can be simplified slightly to obtain,
\begin{equation}\label{eq:BerryConnection}
\begin{split}
\mathbf{A}_{n}(\mathbf{k})=&-i\sum_\mathbf{g\mathbf{g}'}f^\dagger_{n\mathbf{k}+\mathbf{g}}\bar{\mathcal{M}}_{\mathbf{g}'-\mathbf{g}}(\omega_{n\mathbf{k}},\mathbf{k}+\mathbf{g})\partial_\mathbf{k}f_{n\mathbf{k}+\mathbf{g}'}\\
&+\sum_\mathbf{g\mathbf{g}'}f^\dagger_{n\mathbf{k}+\mathbf{g}}\pmb{\mathcal{A}}_{\mathbf{g}'-\mathbf{g}}(\omega_{n\mathbf{k}},\mathbf{k}+\mathbf{g})f_{n\mathbf{k}+\mathbf{g}'}.
\end{split}
\end{equation}
The first term gives the Berry connection of the photon, while the second term $\pmb{\mathcal{A}}_{\mathbf{g}}(\omega,\mathbf{k})$ arises solely from the matter oscillations,
\begin{equation}\label{eq:Berry_Oscillators}
\pmb{\mathcal{A}}_{\mathbf{g}}(\omega,\mathbf{k})=-i\sum_{\alpha\mathbf{g}'}\frac{\mathcal{C}^\dagger_{\alpha-\mathbf{g}'}(\mathbf{k}+\mathbf{g}')\partial_\mathbf{k}\mathcal{C}_{\alpha\mathbf{g}-\mathbf{g}'}(\mathbf{k}+\mathbf{g}')}{(\omega-\omega_{\alpha\mathbf{k}+\mathbf{g}'})^2}.
\end{equation}
Due to nonlocality, Eq. (\ref{eq:Berry_Oscillators}) does not generally vanish. This additional contribution to the Berry phase corresponds to vortices in the response function itself - independent of the Berry gauge of the photon. This means the Chern number can be nonzero $\mathfrak{C}_n\neq 0$ even if the winding of electromagnetic field is trivial. However, we will show in the proceeding sections that all symmetry constraints on the Chern number can be established entirely in terms of the photon.

As can be seen from Eq.~(\ref{eq:BerryConnection}), the Berry connection is only defined within the Brillouin zone $\mathbf{A}_{n\mathbf{k}+\mathbf{g}}=\mathbf{A}_{n\mathbf{k}}+\partial_\mathbf{k}\chi_{n\mathbf{k}}$, up to a possible U(1) gauge. Hence, the gauge invariant Berry curvature is periodic $F_{n\mathbf{k}+\mathbf{g}}=F_{n\mathbf{k}}$,
\begin{equation}
F_{n}(\mathbf{k})=\mathbf{\hat{z}}\cdot[\partial_\mathbf{k}\times\mathbf{A}_n(\mathbf{k})].
\end{equation}
The Chern number is found by integrating the Berry curvature over the two-dimensional Brillouin zone,
\begin{equation}\label{eq:Photonic_Chern}
\mathfrak{C}_n=\frac{1}{2\pi}\int_{\textrm{BZ}}F_n(\mathbf{k})d^2\mathbf{k}, \qquad \mathfrak{C}_n\in \mathbb{Z},
\end{equation}
which determines the winding number of the collective light-matter excitations over the torus $\mathbb{T}^2=S^1\times S^1$. Equation~(\ref{eq:Photonic_Chern}) is one of the central results of this paper. An electromagnetic Chern invariant can be found for any 2D crystal and characterizes distinct topological phases of matter $\mathfrak{C}_n\neq 0$.

\begin{figure*}
  \includegraphics[width=0.97\textwidth]{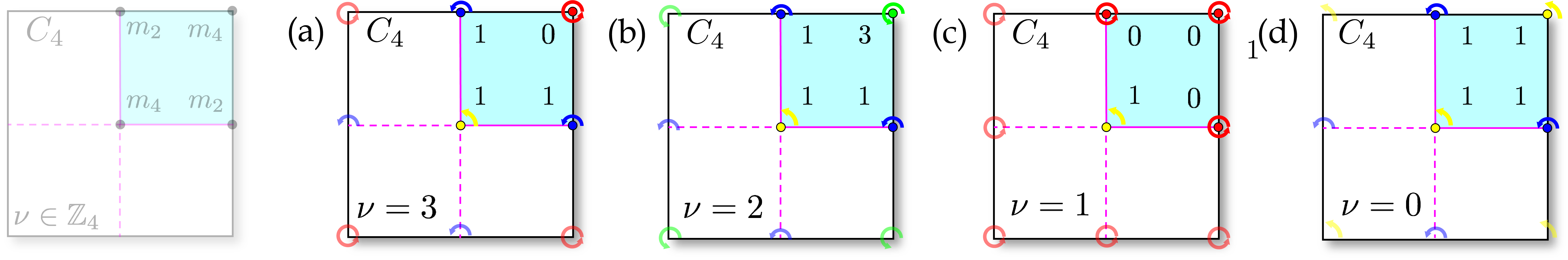}
  \caption{Examples of SPT bosonic phases in a crystal with $C_4$ symmetry. These phases are characterized by their SPT invariant $\nu=m_4(\Gamma)+m_4(\textrm{M})+2m_2(\textrm{Y}) \mod 4$ which determines the electromagnetic Chern number up to a multiple of 4. Here, $m_4\in\mathbb{Z}_4$ and $m_2\in\mathbb{Z}_2$ are modulo integers. (a), (b), (c) and (d) correspond to SPT bosonic phases of $\nu=3,2,1$ and 0 respectively. For bosons, we simply add up all the integer charges within the irreducible Brillouin zone. For instance, the $\nu=2$ phase has eigenvalues of $m_4(\Gamma)=1$ at the center and $m_4(\textrm{M})=3=-1$ at the vertices, with inversion eigenvalues of $m_2(\textrm{Y})=m_2(\textrm{X})=1$ at the edge centers: $\nu=1+3+2\times 1=2$.}
  \label{fig:C4_phases}
    \includegraphics[width=0.97\textwidth]{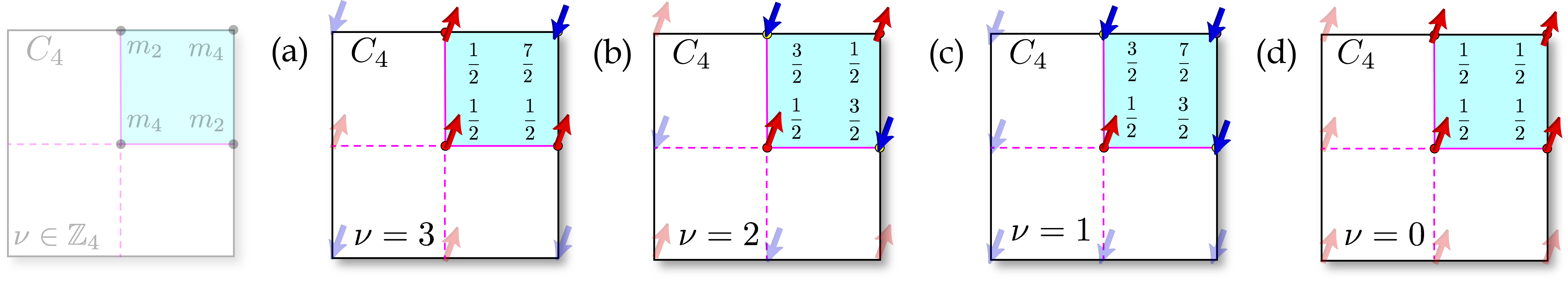}
  \caption{Examples of SPT fermionic phases in a crystal with $C_4$ symmetry. These phases are characterized by their SPT invariant $\nu=m_4(\Gamma)+m_4(\textrm{M})+2m_2(\textrm{Y})+2 \mod 4$ which determines the electronic Chern number up to a multiple of 4. In this case, $m_4\in\mathbb{Z}_4+\sfrac{1}{2}$ and $m_2\in\mathbb{Z}_2+\sfrac{1}{2}$ are modulo half-integers. (a), (b), (c) and (d) correspond to SPT fermionic phases of $\nu=3,2,1$ and 0 respectively. The problem is more complicated for fermions because the charges are fractional and we must also account for the antisymmetric phases of a spinor wave function. As an example, the $\nu=2$ phase has eigenvalues of $m_4(\Gamma)=m_4(\textrm{M})=\sfrac{1}{2}$ at the center and vertices, with inversion eigenvalues of $m_2(\textrm{Y})=m_2(\textrm{X})=\sfrac{3}{2}=-\sfrac{1}{2}$ at the edge centers: $\nu=\sfrac{1}{2}+\sfrac{1}{2}+2\times\sfrac{3}{2}+2=2$.}
    \label{fig:C4_phases_fermion}
\end{figure*}

\subsection{Symmetry-protected topological bosonic phases}

Nevertheless, even if we knew the specifics of the material, evaluating the Chern number by brute force would be a herculean task. Instead, we invoke constraints of the point groups, which constitute a type of \textit{symmetry-protected topological} (SPT) phase \cite{Fu_Liang2011,Hughes2011,Slager2012,Fang2012,Kruthoff2017,Bradlyn2017,Sedrakyan2017,Song2017,Po2017,Thorngren2018,Matsugatani2018,Watanabet2018}. SPT phases are protected by the $N$-fold rotational symmetry of $C_N$ and this gives rise to an additional topological invariant $\nu_n\in\mathbb{Z}_N$. Remarkably, $\nu_n$ is classified entirely from $\eta_n(\mathbf{k}_i)$ eigenvalues at HSPs and requires no complicated integration to compute. This invariant is related to the Chern number up to a multiple of $N$,
\begin{equation}
\nu_n=\mathfrak{C}_n\mod N, \qquad \mathfrak{C}_n\in N\mathbb{Z}+\nu_n.
\end{equation}
The interpretation of $\nu_n$ is quite simple - it tells us the geometric phase around the irreducible Brillouin zone ($\textrm{IBZ}$) of the crystal,
\begin{equation}
\begin{split}
 \exp\left(i\frac{2\pi}{N}\mathfrak{C}_n\right)&=\exp\left(i\int_{\textrm{IBZ}}F_n(\mathbf{k}) d^2\mathbf{k}\right)\\
 &=\exp\left(i\oint_{\partial  \textrm{IBZ}}\mathbf{A}_n(\mathbf{k})\cdot d\mathbf{k}\right),
\end{split}
\end{equation}
where $\partial \textrm{IBZ}$ is the path around $\textrm{IBZ}$. This follows from rotational symmetry of the Berry curvature $F_n(\mathbf{k})=F_n(\mathcal{R}\mathbf{k})$. For instance, the path in $C_4$ is $\partial(\textrm{IBZ})_4=\Gamma\textrm{XMY}\Gamma$. Applying the logarithm, $\nu_n$ is equivalent to,
\begin{equation}
\nu_n=\frac{N}{2\pi}\oint_{\partial\textrm{RBZ}}\mathbf{A}_n(\mathbf{k})\cdot d\mathbf{k}\mod N.
\end{equation}
As we will see more explicitly, $\nu_n$ is tied entirely to $\eta_n$. The reason is subtle - any vortex within the interior of the $\textrm{IBZ}$ contributes a Berry phase of $2\pi$, and by symmetry, there are $N$ such vortices within the total Brillouin zone $\mathfrak{C}_n\to \mathfrak{C}_n+N$. However, this has no effect on $\nu_n\to \nu_n$. Only the vortices lying at HSPs contribute to $\nu_n$ because these come in fractions of $2\pi$.

In the following sections we will discuss the bosonic classification of $\nu_n$ for each cyclic point group and the SPT phases associated with them. We do not present the full derivations here since the rigorous proofs have been carried out by others (see Ref.~\cite{Fang2012}) - we simply state the salient results. For completeness, in Appendix \ref{app:SPTFermionic} we also discuss the SPT \textit{fermionic} phases associated with each point group. We do this to emphasize that fermionic and bosonic systems represent distinct topological field theories, with fundamentally different interpretations. These differences are highlighted with a few examples [Fig.~\ref{fig:C4_phases} and \ref{fig:C4_phases_fermion}].

\subsection{Twofold (inversion) symmetry: $C_2$}

For the $C_2$ point group, or simply inversion symmetry, the SPT phase is related to the Chern number by $\nu_n=\mathfrak{C}_n\mod 2$ which is a $\mathbb{Z}_2$ invariant. There is only one nontrivial SPT phase and it can be found modulo 2 from,
\begin{equation}
\exp\left(i\frac{2\pi}{2}\mathfrak{C}_n\right)=\eta_{2,n}(\Gamma)\eta_{2,n}(\textrm{X})\eta_{2,n}(\textrm{Y})\eta_{2,n}(\textrm{M}).
\end{equation}
Applying the logarithm, this classification can be expressed equivalently in terms of $m_{2,n}\in \mathbb{Z}_2$ inversion eigenvalues, 
\begin{equation}
\nu_n=m_{2,n}(\Gamma)+m_{2,n}(\textrm{X})+m_{2,n}(\textrm{Y})+m_{2,n}(\textrm{M}) \mod 2.
\end{equation}
If the summation of $m_{2,n}$ eigenvalues is odd, the SPT phase is nontrivial $\nu_n=1$ and corresponds to an odd-valued Chern number. Likewise, $\nu_n=0$ is an even-valued Chern number.

\subsection{Threefold symmetry: $C_3$}

$C_3$ is unique because it is the only point group with an odd rotational symmetry - i.e. it lacks inversion symmetry. This means the parity of Chern number (odd or even) is not restricted by the symmetries of the crystal. For $C_3$, the SPT phase is $\nu_n= \mathfrak{C}_n\mod 3$ which is a $\mathbb{Z}_3$ invariant. There are two nontrivial SPT phases and they can be found modulo 3 from,
\begin{equation}
\exp\left(i\frac{2\pi}{3}\mathfrak{C}_n\right)=\eta_{3,n}\left(\Gamma\right)\eta_{3,n}\left(\textrm{K}\right)\eta_{3,n}\left(\textrm{K}'\right).
\end{equation}
This classification is expressed equivalently in terms of quantized modulo 3 integers $m_{3,n}\in \mathbb{Z}_3$ at HSPs,
\begin{equation}
\nu_n=m_{3,n}(\Gamma)+m_{3,n}(\textrm{K})+m_{3,n}(\textrm{K}') \mod 3.
\end{equation}
Note though, odd and even phases are not distinct $\nu=-2=1=4$ under modulo 3.

\subsection{Fourfold symmetry: $C_4$}

For the $C_4$ point group, the SPT phase is related to the Chern number by $\nu_n=\mathfrak{C}_n\mod 4$ which is a $\mathbb{Z}_4$ invariant. There are three nontrivial SPT phases and they can be found modulo 4 from,
\begin{equation}\label{eq:C_4Classification}
\exp\left(i\frac{2\pi}{4}\mathfrak{C}_n\right)=\eta_{4,n}\left(\Gamma\right)\eta_{4,n}\left(\textrm{M}\right)\eta_{2,n}\left(\textrm{Y}\right).
\end{equation}
The classification is expressed equivalently in terms of spin-1 eigenvalues,
\begin{equation}\label{eq:C_4Sum}
\nu_n=m_{4,n}(\Gamma)+m_{4,n}(\textrm{M})+2m_{2,n}(\textrm{Y}) \mod 4,
\end{equation}
where $m_{4,n}(\Gamma)$ \& $m_{4,n}(\textrm{M})\in \mathbb{Z}_4$ are modulo 4 integers and $m_{2,n}(\textrm{Y})\in \mathbb{Z}_2$ is a modulo 2 integer. Examples of all SPT phases of the $C_4$ point group are displayed in Fig.~\ref{fig:C4_phases} and these are compared with their fermionic counterparts in Fig.~\ref{fig:C4_phases_fermion}.

\subsection{Sixfold symmetry: $C_6$}

For the $C_6$ point group, the SPT phase is $\nu_n=\mathfrak{C}_n\mod 6$ which is a $\mathbb{Z}_6$ invariant. There are five nontrivial SPT phases and they can be found modulo 6 from,
\begin{equation}\label{eq:C_6Classification}
\exp\left(i\frac{2\pi}{6}\mathfrak{C}_n\right)=\eta_{6,n}\left(\Gamma\right)\eta_{3,n}\left(\textrm{K}\right)\eta_{2,n}\left(\textrm{M}\right).
\end{equation}
This is equivalent to the summation of spin-1 eigenvalues at the HSPs,
\begin{equation}
\nu_n=m_{6,n}(\Gamma)+2m_{3,n}(\textrm{K})+3m_{2,n}(\textrm{M}) \mod 6,
\end{equation}
where $m_{6,n}(\Gamma)\in \mathbb{Z}_6$ is a modulo 4 integer, $m_{6,n}(\textrm{K})\in \mathbb{Z}_3$ is a modulo 3 integer and $m_{2,n}(\textrm{M})\in \mathbb{Z}_2$ is a modulo 2 integer. This completes the classification of all 2+1D topological electromagnetic (bosonic) phases of matter which is summarized in Tbl.~\ref{tab:SPTPhases}. These are compared alongside their fermionic counterparts in Tbl.~\ref{tab:SPTPhasesFermion}.

\subsection{Continuous symmetry: $C_\infty$}

To finish, we briefly discuss the continuum limit $\mathbf{g}=0$ and the topological phases that can be described by a long wavelength theory $k\approx 0$. The physics is significantly more tractable here and exactly solvable models are possible \cite{VanMechelen2018,van_mechelen_photonic_2018}. In this limit, the rotational symmetry of the crystal is approximately continuous $C_\infty$. The SPT invariant $\nu_n$ and Chern number $\mathfrak{C}_n$ are thus equivalent,
\begin{equation}\label{eq:Continuous_Chern}
\nu_n=\mathfrak{C}_n=m_n(0)-m_n(\infty).
\end{equation}
Note that $\nu_n\in\mathbb{Z}$ and $m_n\in \mathbb{Z}$ are \textit{not} modulo integers in this limit and do not have the same interpretation as the lattice theory. This is because we have gained the full rotational symmetry in the continuum approximation. Clearly though, the eigenvalues must change at HSPs $m_n(0)\neq m_n(\infty)$ for a nontrivial phase to exist $\mathfrak{C}_n\neq 0$. In the continuum regularization, $k_i=0$ represents the $\Gamma$ point and $k_i=\infty$ is interpreted as mapping the vertices of the Brillouin zone into one another.

\section{Conclusions}\label{sec:Conclusions}

In summary, we have developed the complete 2+1D lattice field theory describing all symmetry-protected topological bosonic phases of the photon. To accomplish this, we analyzed the electromagnetic Bloch waves in microscopic crystals and derived the Chern invariant of these light-matter excitations. Thereafter, the rotational symmetries of the crystal were examined extensively and the implications these have on photonic spin. We have studied all two dimensional point groups $C_N$ with nonvanishing Chern number $\mathfrak{C}\neq 0$ and linked the topological invariants directly to spin-1 quantized eigenvalues of the electromagnetic field - establishing the bosonic classification for each topological phase.

\section*{Acknowledgements}

This research was supported by the Defense Advanced Research Projects Agency (DARPA) Nascent Light-Matter Interactions (NLM) Program and the National Science Foundation (NSF) [Grant No. EFMA-1641101].

\appendix

\section*{Appendix}

\section{Symmetry-protected topological fermionic phases}\label{app:SPTFermionic}

For completeness, we examine the SPT fermionic phases associated with each point group $C_N$ and highlight their essential differences from bosons. The most important distinction is how they transform under rotations; half-integer particles are antisymmetric $\mathcal{R}(2\pi)=-1$. In terms of discrete rotations $\hat{\mathcal{R}}_N$ about the $z$-axis, the eigenstates of a Bloch spinor particle satisfy,
\begin{equation}
\hat{\mathcal{R}}_N|\Psi(\mathbf{k}_i)\rangle=\zeta_N(\mathbf{k}_i)|\Psi(\mathbf{k}_i)\rangle,
\end{equation}
where the eigenvalues at HSPs are related by,
\begin{equation}
\zeta_N(\mathbf{k}_i)=\exp\left[i\frac{2\pi}{N} m_N(\mathbf{k}_i)\right],\qquad (\zeta_N)^N=-1.
\end{equation}
$m_N(\mathbf{k}_i)\in \mathbb{Z}_N+\sfrac{1}{2}$ is a modulo half-integer and labels the $N$ possible spin-\sfrac{1}{2} eigenvalues. Notice that $\zeta_N$ represents the $N$th roots of \textit{negative} unity which is characteristic of a fermionic field.

The single-particle fermionic classification for $C_2$, $C_3$, $C_4$ and $C_6$ respectively is \cite{Fang2012,Matsugatani2018},
\begin{subequations}
\begin{equation}
\exp\left(i\frac{2\pi}{2}\mathfrak{C}\right)=\zeta_{2}(\Gamma)\zeta_{2}(\textrm{X})\zeta_{2}(\textrm{Y})\zeta_{2}(\textrm{M}),
\end{equation}
\begin{equation}
\exp\left(i\frac{2\pi}{3}\mathfrak{C}\right)=-\zeta_{3}\left(\Gamma\right)\zeta_{3}\left(\textrm{K}\right)\zeta_{3}\left(\textrm{K}'\right),
\end{equation}
\begin{equation}
\exp\left(i\frac{2\pi}{4}\mathfrak{C}\right)=-\zeta_{4}\left(\Gamma\right)\zeta_{4}\left(\textrm{M}\right)\zeta_{2}\left(\textrm{Y}\right),
\end{equation}
\begin{equation}
\exp\left(i\frac{2\pi}{6}\mathfrak{C}\right)=-\zeta_{6}\left(\Gamma\right)\zeta_{3}\left(\textrm{K}\right)\zeta_{2}\left(\textrm{M}\right).
\end{equation}
\end{subequations}
Although the classification appears similar, the SPT fermionic phases constitute very different physics than their bosonic counterparts, which is alluded to by the antisymmetric phase factors $\mathcal{R}(2\pi)=-1$. We illustrate this with an example in $C_4$. Applying the logarithm - the classification for the SPT fermionic phase $\nu=\mathfrak{C}\mod 4$ can be expressed as,
\begin{equation}
\nu=m_4(\Gamma)+m_4(\textrm{M})+2m_2(\textrm{Y})+2\mod 4,
\end{equation}
where $m_4(\Gamma)$ \& $m_4(\textrm{M})\in \mathbb{Z}_4+\sfrac{1}{2}$ are modulo $4$ half-integers and $m_2(\textrm{Y})\in \mathbb{Z}_2+\sfrac{1}{2}$ is a modulo $2$ half-integer.

\bibliography{spin_quant.bib}

\begin{thebibliography}{93}%
\makeatletter
\providecommand \@ifxundefined [1]{%
 \@ifx{#1\undefined}
}%
\providecommand \@ifnum [1]{%
 \ifnum #1\expandafter \@firstoftwo
 \else \expandafter \@secondoftwo
 \fi
}%
\providecommand \@ifx [1]{%
 \ifx #1\expandafter \@firstoftwo
 \else \expandafter \@secondoftwo
 \fi
}%
\providecommand \natexlab [1]{#1}%
\providecommand \enquote  [1]{``#1''}%
\providecommand \bibnamefont  [1]{#1}%
\providecommand \bibfnamefont [1]{#1}%
\providecommand \citenamefont [1]{#1}%
\providecommand \href@noop [0]{\@secondoftwo}%
\providecommand \href [0]{\begingroup \@sanitize@url \@href}%
\providecommand \@href[1]{\@@startlink{#1}\@@href}%
\providecommand \@@href[1]{\endgroup#1\@@endlink}%
\providecommand \@sanitize@url [0]{\catcode `\\12\catcode `\$12\catcode
  `\&12\catcode `\#12\catcode `\^12\catcode `\_12\catcode `\%12\relax}%
\providecommand \@@startlink[1]{}%
\providecommand \@@endlink[0]{}%
\providecommand \url  [0]{\begingroup\@sanitize@url \@url }%
\providecommand \@url [1]{\endgroup\@href {#1}{\urlprefix }}%
\providecommand \urlprefix  [0]{URL }%
\providecommand \Eprint [0]{\href }%
\providecommand \doibase [0]{http://dx.doi.org/}%
\providecommand \selectlanguage [0]{\@gobble}%
\providecommand \bibinfo  [0]{\@secondoftwo}%
\providecommand \bibfield  [0]{\@secondoftwo}%
\providecommand \translation [1]{[#1]}%
\providecommand \BibitemOpen [0]{}%
\providecommand \bibitemStop [0]{}%
\providecommand \bibitemNoStop [0]{.\EOS\space}%
\providecommand \EOS [0]{\spacefactor3000\relax}%
\providecommand \BibitemShut  [1]{\csname bibitem#1\endcsname}%
\let\auto@bib@innerbib\@empty
\bibitem [{\citenamefont {Hasan}\ and\ \citenamefont {Kane}(2010)}]{Kane2010}%
  \BibitemOpen
  \bibfield  {author} {\bibinfo {author} {\bibfnamefont {M.~Z.}\ \bibnamefont
  {Hasan}}\ and\ \bibinfo {author} {\bibfnamefont {C.~L.}\ \bibnamefont
  {Kane}},\ }\bibfield  {title} {\enquote {\bibinfo {title} {Colloquium:
  Topological insulators},}\ }\href {\doibase 10.1103/RevModPhys.82.3045}
  {\bibfield  {journal} {\bibinfo  {journal} {Rev. Mod. Phys.}\ }\textbf
  {\bibinfo {volume} {82}},\ \bibinfo {pages} {3045--3067} (\bibinfo {year}
  {2010})}\BibitemShut {NoStop}%
\bibitem [{\citenamefont {Ryu}\ \emph {et~al.}(2010)\citenamefont {Ryu},
  \citenamefont {Schnyder}, \citenamefont {Furusaki},\ and\ \citenamefont
  {Ludwig}}]{Ryu2010}%
  \BibitemOpen
  \bibfield  {author} {\bibinfo {author} {\bibfnamefont {Shinsei}\ \bibnamefont
  {Ryu}}, \bibinfo {author} {\bibfnamefont {Andreas~P}\ \bibnamefont
  {Schnyder}}, \bibinfo {author} {\bibfnamefont {Akira}\ \bibnamefont
  {Furusaki}}, \ and\ \bibinfo {author} {\bibfnamefont {Andreas W~W}\
  \bibnamefont {Ludwig}},\ }\bibfield  {title} {\enquote {\bibinfo {title}
  {Topological insulators and superconductors: tenfold way and dimensional
  hierarchy},}\ }\href {http://stacks.iop.org/1367-2630/12/i=6/a=065010}
  {\bibfield  {journal} {\bibinfo  {journal} {New Journal of Physics}\ }\textbf
  {\bibinfo {volume} {12}},\ \bibinfo {pages} {065010} (\bibinfo {year}
  {2010})}\BibitemShut {NoStop}%
\bibitem [{\citenamefont {Haldane}(1988)}]{Haldane1988}%
  \BibitemOpen
  \bibfield  {author} {\bibinfo {author} {\bibfnamefont {F.~D.~M.}\
  \bibnamefont {Haldane}},\ }\bibfield  {title} {\enquote {\bibinfo {title}
  {Model for a quantum hall effect without landau levels: Condensed-matter
  realization of the "parity anomaly"},}\ }\href {\doibase
  10.1103/PhysRevLett.61.2015} {\bibfield  {journal} {\bibinfo  {journal}
  {Phys. Rev. Lett.}\ }\textbf {\bibinfo {volume} {61}},\ \bibinfo {pages}
  {2015--2018} (\bibinfo {year} {1988})}\BibitemShut {NoStop}%
\bibitem [{\citenamefont {Jaksch}\ and\ \citenamefont
  {Zoller}(2003)}]{Zoller2003}%
  \BibitemOpen
  \bibfield  {author} {\bibinfo {author} {\bibfnamefont {D}~\bibnamefont
  {Jaksch}}\ and\ \bibinfo {author} {\bibfnamefont {P}~\bibnamefont {Zoller}},\
  }\bibfield  {title} {\enquote {\bibinfo {title} {Creation of effective
  magnetic fields in optical lattices: the hofstadter butterfly for cold
  neutral atoms},}\ }\href {http://stacks.iop.org/1367-2630/5/i=1/a=356}
  {\bibfield  {journal} {\bibinfo  {journal} {New Journal of Physics}\ }\textbf
  {\bibinfo {volume} {5}},\ \bibinfo {pages} {56} (\bibinfo {year}
  {2003})}\BibitemShut {NoStop}%
\bibitem [{\citenamefont {Koch}\ \emph {et~al.}(2010)\citenamefont {Koch},
  \citenamefont {Houck}, \citenamefont {Hur},\ and\ \citenamefont
  {Girvin}}]{Houck2010}%
  \BibitemOpen
  \bibfield  {author} {\bibinfo {author} {\bibfnamefont {Jens}\ \bibnamefont
  {Koch}}, \bibinfo {author} {\bibfnamefont {Andrew~A.}\ \bibnamefont {Houck}},
  \bibinfo {author} {\bibfnamefont {Karyn~Le}\ \bibnamefont {Hur}}, \ and\
  \bibinfo {author} {\bibfnamefont {S.~M.}\ \bibnamefont {Girvin}},\ }\bibfield
   {title} {\enquote {\bibinfo {title} {Time-reversal-symmetry breaking in
  circuit-qed-based photon lattices},}\ }\href {\doibase
  10.1103/PhysRevA.82.043811} {\bibfield  {journal} {\bibinfo  {journal} {Phys.
  Rev. A}\ }\textbf {\bibinfo {volume} {82}},\ \bibinfo {pages} {043811}
  (\bibinfo {year} {2010})}\BibitemShut {NoStop}%
\bibitem [{\citenamefont {Jotzu}\ \emph {et~al.}(2014)\citenamefont {Jotzu},
  \citenamefont {Messer}, \citenamefont {Desbuquois}, \citenamefont {Lebrat},
  \citenamefont {Uehlinger}, \citenamefont {Greif},\ and\ \citenamefont
  {Esslinger}}]{Jotzu2014}%
  \BibitemOpen
  \bibfield  {author} {\bibinfo {author} {\bibfnamefont {Gregor}\ \bibnamefont
  {Jotzu}}, \bibinfo {author} {\bibfnamefont {Michael}\ \bibnamefont {Messer}},
  \bibinfo {author} {\bibfnamefont {R{\'e}mi}\ \bibnamefont {Desbuquois}},
  \bibinfo {author} {\bibfnamefont {Martin}\ \bibnamefont {Lebrat}}, \bibinfo
  {author} {\bibfnamefont {Thomas}\ \bibnamefont {Uehlinger}}, \bibinfo
  {author} {\bibfnamefont {Daniel}\ \bibnamefont {Greif}}, \ and\ \bibinfo
  {author} {\bibfnamefont {Tilman}\ \bibnamefont {Esslinger}},\ }\bibfield
  {title} {\enquote {\bibinfo {title} {Experimental realization of the
  topological haldane model with ultracold fermions},}\ }\href
  {http://dx.doi.org/10.1038/nature13915} {\bibfield  {journal} {\bibinfo
  {journal} {Nature}\ }\textbf {\bibinfo {volume} {515}},\ \bibinfo {pages}
  {237 EP --} (\bibinfo {year} {2014})}\BibitemShut {NoStop}%
\bibitem [{\citenamefont {Kane}\ and\ \citenamefont {Mele}(2005)}]{Kane2005}%
  \BibitemOpen
  \bibfield  {author} {\bibinfo {author} {\bibfnamefont {C.~L.}\ \bibnamefont
  {Kane}}\ and\ \bibinfo {author} {\bibfnamefont {E.~J.}\ \bibnamefont
  {Mele}},\ }\bibfield  {title} {\enquote {\bibinfo {title} {Quantum spin hall
  effect in graphene},}\ }\href {\doibase 10.1103/PhysRevLett.95.226801}
  {\bibfield  {journal} {\bibinfo  {journal} {Phys. Rev. Lett.}\ }\textbf
  {\bibinfo {volume} {95}},\ \bibinfo {pages} {226801} (\bibinfo {year}
  {2005})}\BibitemShut {NoStop}%
\bibitem [{\citenamefont {Bernevig}\ \emph {et~al.}(2006)\citenamefont
  {Bernevig}, \citenamefont {Hughes},\ and\ \citenamefont
  {Zhang}}]{Bernevig1757}%
  \BibitemOpen
  \bibfield  {author} {\bibinfo {author} {\bibfnamefont {B.~Andrei}\
  \bibnamefont {Bernevig}}, \bibinfo {author} {\bibfnamefont {Taylor~L.}\
  \bibnamefont {Hughes}}, \ and\ \bibinfo {author} {\bibfnamefont {Shou-Cheng}\
  \bibnamefont {Zhang}},\ }\bibfield  {title} {\enquote {\bibinfo {title}
  {Quantum spin hall effect and topological phase transition in hgte quantum
  wells},}\ }\href {\doibase 10.1126/science.1133734} {\bibfield  {journal}
  {\bibinfo  {journal} {Science}\ }\textbf {\bibinfo {volume} {314}},\ \bibinfo
  {pages} {1757--1761} (\bibinfo {year} {2006})}\BibitemShut {NoStop}%
\bibitem [{\citenamefont {Maciejko}\ \emph {et~al.}(2011)\citenamefont
  {Maciejko}, \citenamefont {Hughes},\ and\ \citenamefont
  {Zhang}}]{Maciejko2011}%
  \BibitemOpen
  \bibfield  {author} {\bibinfo {author} {\bibfnamefont {Joseph}\ \bibnamefont
  {Maciejko}}, \bibinfo {author} {\bibfnamefont {Taylor~L.}\ \bibnamefont
  {Hughes}}, \ and\ \bibinfo {author} {\bibfnamefont {Shou-Cheng}\ \bibnamefont
  {Zhang}},\ }\bibfield  {title} {\enquote {\bibinfo {title} {The quantum spin
  hall effect},}\ }\href {\doibase 10.1146/annurev-conmatphys-062910-140538}
  {\bibfield  {journal} {\bibinfo  {journal} {Annual Review of Condensed Matter
  Physics}\ }\textbf {\bibinfo {volume} {2}},\ \bibinfo {pages} {31--53}
  (\bibinfo {year} {2011})}\BibitemShut {NoStop}%
\bibitem [{\citenamefont {Klitzing}\ \emph {et~al.}(1980)\citenamefont
  {Klitzing}, \citenamefont {Dorda},\ and\ \citenamefont
  {Pepper}}]{Klitzing1980}%
  \BibitemOpen
  \bibfield  {author} {\bibinfo {author} {\bibfnamefont {K.~v.}\ \bibnamefont
  {Klitzing}}, \bibinfo {author} {\bibfnamefont {G.}~\bibnamefont {Dorda}}, \
  and\ \bibinfo {author} {\bibfnamefont {M.}~\bibnamefont {Pepper}},\
  }\bibfield  {title} {\enquote {\bibinfo {title} {New method for high-accuracy
  determination of the fine-structure constant based on quantized hall
  resistance},}\ }\href {\doibase 10.1103/PhysRevLett.45.494} {\bibfield
  {journal} {\bibinfo  {journal} {Phys. Rev. Lett.}\ }\textbf {\bibinfo
  {volume} {45}},\ \bibinfo {pages} {494--497} (\bibinfo {year}
  {1980})}\BibitemShut {NoStop}%
\bibitem [{\citenamefont {Laughlin}(1981)}]{Laughlin1981}%
  \BibitemOpen
  \bibfield  {author} {\bibinfo {author} {\bibfnamefont {R.~B.}\ \bibnamefont
  {Laughlin}},\ }\bibfield  {title} {\enquote {\bibinfo {title} {Quantized hall
  conductivity in two dimensions},}\ }\href {\doibase 10.1103/PhysRevB.23.5632}
  {\bibfield  {journal} {\bibinfo  {journal} {Phys. Rev. B}\ }\textbf {\bibinfo
  {volume} {23}},\ \bibinfo {pages} {5632--5633} (\bibinfo {year}
  {1981})}\BibitemShut {NoStop}%
\bibitem [{\citenamefont {Thouless}\ \emph {et~al.}(1982)\citenamefont
  {Thouless}, \citenamefont {Kohmoto}, \citenamefont {Nightingale},\ and\
  \citenamefont {den Nijs}}]{Thouless1982}%
  \BibitemOpen
  \bibfield  {author} {\bibinfo {author} {\bibfnamefont {D.~J.}\ \bibnamefont
  {Thouless}}, \bibinfo {author} {\bibfnamefont {M.}~\bibnamefont {Kohmoto}},
  \bibinfo {author} {\bibfnamefont {M.~P.}\ \bibnamefont {Nightingale}}, \ and\
  \bibinfo {author} {\bibfnamefont {M.}~\bibnamefont {den Nijs}},\ }\bibfield
  {title} {\enquote {\bibinfo {title} {Quantized hall conductance in a
  two-dimensional periodic potential},}\ }\href {\doibase
  10.1103/PhysRevLett.49.405} {\bibfield  {journal} {\bibinfo  {journal} {Phys.
  Rev. Lett.}\ }\textbf {\bibinfo {volume} {49}},\ \bibinfo {pages} {405--408}
  (\bibinfo {year} {1982})}\BibitemShut {NoStop}%
\bibitem [{\citenamefont {Chen}\ \emph {et~al.}(2011)\citenamefont {Chen},
  \citenamefont {Liu},\ and\ \citenamefont {Wen}}]{Wen2011}%
  \BibitemOpen
  \bibfield  {author} {\bibinfo {author} {\bibfnamefont {Xie}\ \bibnamefont
  {Chen}}, \bibinfo {author} {\bibfnamefont {Zheng-Xin}\ \bibnamefont {Liu}}, \
  and\ \bibinfo {author} {\bibfnamefont {Xiao-Gang}\ \bibnamefont {Wen}},\
  }\bibfield  {title} {\enquote {\bibinfo {title} {Two-dimensional
  symmetry-protected topological orders and their protected gapless edge
  excitations},}\ }\href {\doibase 10.1103/PhysRevB.84.235141} {\bibfield
  {journal} {\bibinfo  {journal} {Phys. Rev. B}\ }\textbf {\bibinfo {volume}
  {84}},\ \bibinfo {pages} {235141} (\bibinfo {year} {2011})}\BibitemShut
  {NoStop}%
\bibitem [{\citenamefont {Chen}\ \emph {et~al.}(2012)\citenamefont {Chen},
  \citenamefont {Gu}, \citenamefont {Liu},\ and\ \citenamefont
  {Wen}}]{Chen1604}%
  \BibitemOpen
  \bibfield  {author} {\bibinfo {author} {\bibfnamefont {Xie}\ \bibnamefont
  {Chen}}, \bibinfo {author} {\bibfnamefont {Zheng-Cheng}\ \bibnamefont {Gu}},
  \bibinfo {author} {\bibfnamefont {Zheng-Xin}\ \bibnamefont {Liu}}, \ and\
  \bibinfo {author} {\bibfnamefont {Xiao-Gang}\ \bibnamefont {Wen}},\
  }\bibfield  {title} {\enquote {\bibinfo {title} {Symmetry-protected
  topological orders in interacting bosonic systems},}\ }\href {\doibase
  10.1126/science.1227224} {\bibfield  {journal} {\bibinfo  {journal}
  {Science}\ }\textbf {\bibinfo {volume} {338}},\ \bibinfo {pages} {1604--1606}
  (\bibinfo {year} {2012})}\BibitemShut {NoStop}%
\bibitem [{\citenamefont {Lu}\ and\ \citenamefont
  {Vishwanath}(2012)}]{Vishwanath2012}%
  \BibitemOpen
  \bibfield  {author} {\bibinfo {author} {\bibfnamefont {Yuan-Ming}\
  \bibnamefont {Lu}}\ and\ \bibinfo {author} {\bibfnamefont {Ashvin}\
  \bibnamefont {Vishwanath}},\ }\bibfield  {title} {\enquote {\bibinfo {title}
  {Theory and classification of interacting integer topological phases in two
  dimensions: A chern-simons approach},}\ }\href {\doibase
  10.1103/PhysRevB.86.125119} {\bibfield  {journal} {\bibinfo  {journal} {Phys.
  Rev. B}\ }\textbf {\bibinfo {volume} {86}},\ \bibinfo {pages} {125119}
  (\bibinfo {year} {2012})}\BibitemShut {NoStop}%
\bibitem [{\citenamefont {Vishwanath}\ and\ \citenamefont
  {Senthil}(2013)}]{Vishwanath2013}%
  \BibitemOpen
  \bibfield  {author} {\bibinfo {author} {\bibfnamefont {Ashvin}\ \bibnamefont
  {Vishwanath}}\ and\ \bibinfo {author} {\bibfnamefont {T.}~\bibnamefont
  {Senthil}},\ }\bibfield  {title} {\enquote {\bibinfo {title} {Physics of
  three-dimensional bosonic topological insulators: Surface-deconfined
  criticality and quantized magnetoelectric effect},}\ }\href {\doibase
  10.1103/PhysRevX.3.011016} {\bibfield  {journal} {\bibinfo  {journal} {Phys.
  Rev. X}\ }\textbf {\bibinfo {volume} {3}},\ \bibinfo {pages} {011016}
  (\bibinfo {year} {2013})}\BibitemShut {NoStop}%
\bibitem [{\citenamefont {Metlitski}\ \emph {et~al.}(2013)\citenamefont
  {Metlitski}, \citenamefont {Kane},\ and\ \citenamefont
  {Fisher}}]{Metlitski2013}%
  \BibitemOpen
  \bibfield  {author} {\bibinfo {author} {\bibfnamefont {Max~A.}\ \bibnamefont
  {Metlitski}}, \bibinfo {author} {\bibfnamefont {C.~L.}\ \bibnamefont {Kane}},
  \ and\ \bibinfo {author} {\bibfnamefont {Matthew P.~A.}\ \bibnamefont
  {Fisher}},\ }\bibfield  {title} {\enquote {\bibinfo {title} {Bosonic
  topological insulator in three dimensions and the statistical witten
  effect},}\ }\href {\doibase 10.1103/PhysRevB.88.035131} {\bibfield  {journal}
  {\bibinfo  {journal} {Phys. Rev. B}\ }\textbf {\bibinfo {volume} {88}},\
  \bibinfo {pages} {035131} (\bibinfo {year} {2013})}\BibitemShut {NoStop}%
\bibitem [{\citenamefont {Senthil}\ and\ \citenamefont
  {Levin}(2013)}]{Senthil2013}%
  \BibitemOpen
  \bibfield  {author} {\bibinfo {author} {\bibfnamefont {T.}~\bibnamefont
  {Senthil}}\ and\ \bibinfo {author} {\bibfnamefont {Michael}\ \bibnamefont
  {Levin}},\ }\bibfield  {title} {\enquote {\bibinfo {title} {Integer quantum
  hall effect for bosons},}\ }\href {\doibase 10.1103/PhysRevLett.110.046801}
  {\bibfield  {journal} {\bibinfo  {journal} {Phys. Rev. Lett.}\ }\textbf
  {\bibinfo {volume} {110}},\ \bibinfo {pages} {046801} (\bibinfo {year}
  {2013})}\BibitemShut {NoStop}%
\bibitem [{\citenamefont {Lin}\ and\ \citenamefont {Liu}(2015)}]{Lin2015}%
  \BibitemOpen
  \bibfield  {author} {\bibinfo {author} {\bibfnamefont {Zeren}\ \bibnamefont
  {Lin}}\ and\ \bibinfo {author} {\bibfnamefont {Zhirong}\ \bibnamefont
  {Liu}},\ }\bibfield  {title} {\enquote {\bibinfo {title} {Spin-1 dirac-weyl
  fermions protected by bipartite symmetry},}\ }\href {\doibase
  10.1063/1.4936774} {\bibfield  {journal} {\bibinfo  {journal} {The Journal of
  Chemical Physics}\ }\textbf {\bibinfo {volume} {143}},\ \bibinfo {pages}
  {214109} (\bibinfo {year} {2015})}\BibitemShut {NoStop}%
\bibitem [{\citenamefont {Lan}\ \emph {et~al.}(2016)\citenamefont {Lan},
  \citenamefont {Kong},\ and\ \citenamefont {Wen}}]{Tian2016}%
  \BibitemOpen
  \bibfield  {author} {\bibinfo {author} {\bibfnamefont {Tian}\ \bibnamefont
  {Lan}}, \bibinfo {author} {\bibfnamefont {Liang}\ \bibnamefont {Kong}}, \
  and\ \bibinfo {author} {\bibfnamefont {Xiao-Gang}\ \bibnamefont {Wen}},\
  }\bibfield  {title} {\enquote {\bibinfo {title} {Theory of (2+1)-dimensional
  fermionic topological orders and fermionic/bosonic topological orders with
  symmetries},}\ }\href {\doibase 10.1103/PhysRevB.94.155113} {\bibfield
  {journal} {\bibinfo  {journal} {Phys. Rev. B}\ }\textbf {\bibinfo {volume}
  {94}},\ \bibinfo {pages} {155113} (\bibinfo {year} {2016})}\BibitemShut
  {NoStop}%
\bibitem [{\citenamefont {Zhang}\ \emph {et~al.}(2018)\citenamefont {Zhang},
  \citenamefont {Song}, \citenamefont {Alexandradinata}, \citenamefont {Weng},
  \citenamefont {Fang}, \citenamefont {Lu},\ and\ \citenamefont
  {Fang}}]{Zhang2018}%
  \BibitemOpen
  \bibfield  {author} {\bibinfo {author} {\bibfnamefont {Tiantian}\
  \bibnamefont {Zhang}}, \bibinfo {author} {\bibfnamefont {Zhida}\ \bibnamefont
  {Song}}, \bibinfo {author} {\bibfnamefont {A.}~\bibnamefont
  {Alexandradinata}}, \bibinfo {author} {\bibfnamefont {Hongming}\ \bibnamefont
  {Weng}}, \bibinfo {author} {\bibfnamefont {Chen}\ \bibnamefont {Fang}},
  \bibinfo {author} {\bibfnamefont {Ling}\ \bibnamefont {Lu}}, \ and\ \bibinfo
  {author} {\bibfnamefont {Zhong}\ \bibnamefont {Fang}},\ }\bibfield  {title}
  {\enquote {\bibinfo {title} {Double-weyl phonons in transition-metal
  monosilicides},}\ }\href {\doibase 10.1103/PhysRevLett.120.016401} {\bibfield
   {journal} {\bibinfo  {journal} {Phys. Rev. Lett.}\ }\textbf {\bibinfo
  {volume} {120}},\ \bibinfo {pages} {016401} (\bibinfo {year}
  {2018})}\BibitemShut {NoStop}%
\bibitem [{\citenamefont {Ikebe}\ \emph {et~al.}(2010)\citenamefont {Ikebe},
  \citenamefont {Morimoto}, \citenamefont {Masutomi}, \citenamefont {Okamoto},
  \citenamefont {Aoki},\ and\ \citenamefont {Shimano}}]{Shimano2010}%
  \BibitemOpen
  \bibfield  {author} {\bibinfo {author} {\bibfnamefont {Y.}~\bibnamefont
  {Ikebe}}, \bibinfo {author} {\bibfnamefont {T.}~\bibnamefont {Morimoto}},
  \bibinfo {author} {\bibfnamefont {R.}~\bibnamefont {Masutomi}}, \bibinfo
  {author} {\bibfnamefont {T.}~\bibnamefont {Okamoto}}, \bibinfo {author}
  {\bibfnamefont {H.}~\bibnamefont {Aoki}}, \ and\ \bibinfo {author}
  {\bibfnamefont {R.}~\bibnamefont {Shimano}},\ }\bibfield  {title} {\enquote
  {\bibinfo {title} {Optical hall effect in the integer quantum hall regime},}\
  }\href {\doibase 10.1103/PhysRevLett.104.256802} {\bibfield  {journal}
  {\bibinfo  {journal} {Phys. Rev. Lett.}\ }\textbf {\bibinfo {volume} {104}},\
  \bibinfo {pages} {256802} (\bibinfo {year} {2010})}\BibitemShut {NoStop}%
\bibitem [{\citenamefont {Zheng}\ and\ \citenamefont {Ando}(2002)}]{Zheng2002}%
  \BibitemOpen
  \bibfield  {author} {\bibinfo {author} {\bibfnamefont {Yisong}\ \bibnamefont
  {Zheng}}\ and\ \bibinfo {author} {\bibfnamefont {Tsuneya}\ \bibnamefont
  {Ando}},\ }\bibfield  {title} {\enquote {\bibinfo {title} {Hall conductivity
  of a two-dimensional graphite system},}\ }\href {\doibase
  10.1103/PhysRevB.65.245420} {\bibfield  {journal} {\bibinfo  {journal} {Phys.
  Rev. B}\ }\textbf {\bibinfo {volume} {65}},\ \bibinfo {pages} {245420}
  (\bibinfo {year} {2002})}\BibitemShut {NoStop}%
\bibitem [{\citenamefont {Van~Mechelen}\ and\ \citenamefont
  {Jacob}(2018{\natexlab{a}})}]{VanMechelen2018}%
  \BibitemOpen
  \bibfield  {author} {\bibinfo {author} {\bibfnamefont {Todd}\ \bibnamefont
  {Van~Mechelen}}\ and\ \bibinfo {author} {\bibfnamefont {Zubin}\ \bibnamefont
  {Jacob}},\ }\bibfield  {title} {\enquote {\bibinfo {title} {Quantum
  gyroelectric effect: Photon spin-1 quantization in continuum topological
  bosonic phases},}\ }\href {\doibase 10.1103/PhysRevA.98.023842} {\bibfield
  {journal} {\bibinfo  {journal} {Phys. Rev. A}\ }\textbf {\bibinfo {volume}
  {98}},\ \bibinfo {pages} {023842} (\bibinfo {year}
  {2018}{\natexlab{a}})}\BibitemShut {NoStop}%
\bibitem [{\citenamefont {Van~Mechelen}\ and\ \citenamefont
  {Jacob}(2018{\natexlab{b}})}]{van_mechelen_photonic_2018}%
  \BibitemOpen
  \bibfield  {author} {\bibinfo {author} {\bibfnamefont {Todd}\ \bibnamefont
  {Van~Mechelen}}\ and\ \bibinfo {author} {\bibfnamefont {Zubin}\ \bibnamefont
  {Jacob}},\ }\bibfield  {title} {\enquote {\bibinfo {title} {Photonic {Dirac}
  monopoles and skyrmions: spin-1 quantization},}\ }\href
  {https://arxiv.org/abs/1806.09879} {\bibfield  {journal} {\bibinfo  {journal}
  {arXiv:1806.09879}\ } (\bibinfo {year} {2018}{\natexlab{b}})}\BibitemShut
  {NoStop}%
\bibitem [{\citenamefont {Dunne}(1999)}]{Dunne1999}%
  \BibitemOpen
  \bibfield  {author} {\bibinfo {author} {\bibfnamefont {G.~V.}\ \bibnamefont
  {Dunne}},\ }\bibfield  {title} {\enquote {\bibinfo {title} {Aspects of
  chern-simons theory},}\ }in\ \href@noop {} {\emph {\bibinfo {booktitle}
  {Aspects topologiques de la physique en basse dimension. Topological aspects
  of low dimensional systems}}},\ \bibinfo {editor} {edited by\ \bibinfo
  {editor} {\bibfnamefont {A.}~\bibnamefont {Comtet}}, \bibinfo {editor}
  {\bibfnamefont {T.}~\bibnamefont {Jolic{\oe}ur}}, \bibinfo {editor}
  {\bibfnamefont {S.}~\bibnamefont {Ouvry}}, \ and\ \bibinfo {editor}
  {\bibfnamefont {F.}~\bibnamefont {David}}}\ (\bibinfo  {publisher} {Springer
  Berlin Heidelberg},\ \bibinfo {address} {Berlin, Heidelberg},\ \bibinfo
  {year} {1999})\ pp.\ \bibinfo {pages} {177--263}\BibitemShut {NoStop}%
\bibitem [{\citenamefont {Boyanovsky}\ \emph {et~al.}(1986)\citenamefont
  {Boyanovsky}, \citenamefont {Blankenbecler},\ and\ \citenamefont
  {Yahalom}}]{BOYANOVSKY1986483}%
  \BibitemOpen
  \bibfield  {author} {\bibinfo {author} {\bibfnamefont {D.}~\bibnamefont
  {Boyanovsky}}, \bibinfo {author} {\bibfnamefont {R.}~\bibnamefont
  {Blankenbecler}}, \ and\ \bibinfo {author} {\bibfnamefont {R.}~\bibnamefont
  {Yahalom}},\ }\bibfield  {title} {\enquote {\bibinfo {title} {Physical origin
  of topological mass in 2 + 1 dimensions},}\ }\href {\doibase
  https://doi.org/10.1016/0550-3213(86)90564-X} {\bibfield  {journal} {\bibinfo
   {journal} {Nuclear Physics B}\ }\textbf {\bibinfo {volume} {270}},\ \bibinfo
  {pages} {483 -- 505} (\bibinfo {year} {1986})}\BibitemShut {NoStop}%
\bibitem [{\citenamefont {Horsley}(2018)}]{Horsley2018}%
  \BibitemOpen
  \bibfield  {author} {\bibinfo {author} {\bibfnamefont {S.~A.~R.}\
  \bibnamefont {Horsley}},\ }\bibfield  {title} {\enquote {\bibinfo {title}
  {Topology and the optical dirac equation},}\ }\href {\doibase
  10.1103/PhysRevA.98.043837} {\bibfield  {journal} {\bibinfo  {journal} {Phys.
  Rev. A}\ }\textbf {\bibinfo {volume} {98}},\ \bibinfo {pages} {043837}
  (\bibinfo {year} {2018})}\BibitemShut {NoStop}%
\bibitem [{\citenamefont {Raghu}\ and\ \citenamefont
  {Haldane}(2008)}]{Haldane2008}%
  \BibitemOpen
  \bibfield  {author} {\bibinfo {author} {\bibfnamefont {S.}~\bibnamefont
  {Raghu}}\ and\ \bibinfo {author} {\bibfnamefont {F.~D.~M.}\ \bibnamefont
  {Haldane}},\ }\bibfield  {title} {\enquote {\bibinfo {title} {Analogs of
  quantum-hall-effect edge states in photonic crystals},}\ }\href {\doibase
  10.1103/PhysRevA.78.033834} {\bibfield  {journal} {\bibinfo  {journal} {Phys.
  Rev. A}\ }\textbf {\bibinfo {volume} {78}},\ \bibinfo {pages} {033834}
  (\bibinfo {year} {2008})}\BibitemShut {NoStop}%
\bibitem [{\citenamefont {Haldane}\ and\ \citenamefont
  {Raghu}(2008)}]{Haldane2008_2}%
  \BibitemOpen
  \bibfield  {author} {\bibinfo {author} {\bibfnamefont {F.~D.~M.}\
  \bibnamefont {Haldane}}\ and\ \bibinfo {author} {\bibfnamefont
  {S.}~\bibnamefont {Raghu}},\ }\bibfield  {title} {\enquote {\bibinfo {title}
  {Possible realization of directional optical waveguides in photonic crystals
  with broken time-reversal symmetry},}\ }\href {\doibase
  10.1103/PhysRevLett.100.013904} {\bibfield  {journal} {\bibinfo  {journal}
  {Phys. Rev. Lett.}\ }\textbf {\bibinfo {volume} {100}},\ \bibinfo {pages}
  {013904} (\bibinfo {year} {2008})}\BibitemShut {NoStop}%
\bibitem [{\citenamefont {Wang}\ \emph {et~al.}(2009)\citenamefont {Wang},
  \citenamefont {Chong}, \citenamefont {Joannopoulos},\ and\ \citenamefont
  {Soljacic}}]{Wang2009}%
  \BibitemOpen
  \bibfield  {author} {\bibinfo {author} {\bibfnamefont {Zheng}\ \bibnamefont
  {Wang}}, \bibinfo {author} {\bibfnamefont {Yidong}\ \bibnamefont {Chong}},
  \bibinfo {author} {\bibfnamefont {J.~D.}\ \bibnamefont {Joannopoulos}}, \
  and\ \bibinfo {author} {\bibfnamefont {Marin}\ \bibnamefont {Soljacic}},\
  }\bibfield  {title} {\enquote {\bibinfo {title} {Observation of
  unidirectional backscattering-immune topological electromagnetic states},}\
  }\href {\doibase 10.1038/nature08293} {\bibfield  {journal} {\bibinfo
  {journal} {Nature}\ }\textbf {\bibinfo {volume} {461}},\ \bibinfo {pages}
  {772--775} (\bibinfo {year} {2009})}\BibitemShut {NoStop}%
\bibitem [{\citenamefont {Wang}\ and\ \citenamefont {Fan}(2005)}]{Wang:05}%
  \BibitemOpen
  \bibfield  {author} {\bibinfo {author} {\bibfnamefont {Zheng}\ \bibnamefont
  {Wang}}\ and\ \bibinfo {author} {\bibfnamefont {Shanhui}\ \bibnamefont
  {Fan}},\ }\bibfield  {title} {\enquote {\bibinfo {title} {Optical circulators
  in two-dimensional magneto-optical photonic crystals},}\ }\href {\doibase
  10.1364/OL.30.001989} {\bibfield  {journal} {\bibinfo  {journal} {Opt.
  Lett.}\ }\textbf {\bibinfo {volume} {30}},\ \bibinfo {pages} {1989--1991}
  (\bibinfo {year} {2005})}\BibitemShut {NoStop}%
\bibitem [{\citenamefont {Lu}\ \emph {et~al.}(2013)\citenamefont {Lu},
  \citenamefont {Fu}, \citenamefont {Joannopoulos},\ and\ \citenamefont
  {Soljacic}}]{Lu2013}%
  \BibitemOpen
  \bibfield  {author} {\bibinfo {author} {\bibfnamefont {Ling}\ \bibnamefont
  {Lu}}, \bibinfo {author} {\bibfnamefont {Liang}\ \bibnamefont {Fu}}, \bibinfo
  {author} {\bibfnamefont {John~D.}\ \bibnamefont {Joannopoulos}}, \ and\
  \bibinfo {author} {\bibfnamefont {Marin}\ \bibnamefont {Soljacic}},\
  }\bibfield  {title} {\enquote {\bibinfo {title} {Weyl points and line nodes
  in gyroid photonic crystals},}\ }\href {\doibase 10.1038/nphoton.2013.42}
  {\bibfield  {journal} {\bibinfo  {journal} {Nat Photon}\ }\textbf {\bibinfo
  {volume} {7}},\ \bibinfo {pages} {294--299} (\bibinfo {year}
  {2013})}\BibitemShut {NoStop}%
\bibitem [{\citenamefont {Hafezi}\ \emph {et~al.}(2013)\citenamefont {Hafezi},
  \citenamefont {Mittal}, \citenamefont {Fan}, \citenamefont {Migdall},\ and\
  \citenamefont {Taylor}}]{HafeziM.2013}%
  \BibitemOpen
  \bibfield  {author} {\bibinfo {author} {\bibfnamefont {M.}~\bibnamefont
  {Hafezi}}, \bibinfo {author} {\bibfnamefont {S.}~\bibnamefont {Mittal}},
  \bibinfo {author} {\bibfnamefont {J.}~\bibnamefont {Fan}}, \bibinfo {author}
  {\bibfnamefont {A.}~\bibnamefont {Migdall}}, \ and\ \bibinfo {author}
  {\bibfnamefont {J.~M.}\ \bibnamefont {Taylor}},\ }\bibfield  {title}
  {\enquote {\bibinfo {title} {Imaging topological edge states in silicon
  photonics},}\ }\href {http://dx.doi.org/10.1038/nphoton.2013.274} {\bibfield
  {journal} {\bibinfo  {journal} {Nat Photon}\ }\textbf {\bibinfo {volume}
  {7}},\ \bibinfo {pages} {1001--1005} (\bibinfo {year} {2013})},\ \bibinfo
  {note} {article}\BibitemShut {NoStop}%
\bibitem [{\citenamefont {Gu}\ \emph {et~al.}(2015)\citenamefont {Gu},
  \citenamefont {Wang},\ and\ \citenamefont {Wen}}]{Zheng2015}%
  \BibitemOpen
  \bibfield  {author} {\bibinfo {author} {\bibfnamefont {Zheng-Cheng}\
  \bibnamefont {Gu}}, \bibinfo {author} {\bibfnamefont {Zhenghan}\ \bibnamefont
  {Wang}}, \ and\ \bibinfo {author} {\bibfnamefont {Xiao-Gang}\ \bibnamefont
  {Wen}},\ }\bibfield  {title} {\enquote {\bibinfo {title} {Classification of
  two-dimensional fermionic and bosonic topological orders},}\ }\href {\doibase
  10.1103/PhysRevB.91.125149} {\bibfield  {journal} {\bibinfo  {journal} {Phys.
  Rev. B}\ }\textbf {\bibinfo {volume} {91}},\ \bibinfo {pages} {125149}
  (\bibinfo {year} {2015})}\BibitemShut {NoStop}%
\bibitem [{\citenamefont {Karzig}\ \emph {et~al.}(2015)\citenamefont {Karzig},
  \citenamefont {Bardyn}, \citenamefont {Lindner},\ and\ \citenamefont
  {Refael}}]{Karzig2015}%
  \BibitemOpen
  \bibfield  {author} {\bibinfo {author} {\bibfnamefont {Torsten}\ \bibnamefont
  {Karzig}}, \bibinfo {author} {\bibfnamefont {Charles-Edouard}\ \bibnamefont
  {Bardyn}}, \bibinfo {author} {\bibfnamefont {Netanel~H.}\ \bibnamefont
  {Lindner}}, \ and\ \bibinfo {author} {\bibfnamefont {Gil}\ \bibnamefont
  {Refael}},\ }\bibfield  {title} {\enquote {\bibinfo {title} {Topological
  polaritons},}\ }\href {\doibase 10.1103/PhysRevX.5.031001} {\bibfield
  {journal} {\bibinfo  {journal} {Phys. Rev. X}\ }\textbf {\bibinfo {volume}
  {5}},\ \bibinfo {pages} {031001} (\bibinfo {year} {2015})}\BibitemShut
  {NoStop}%
\bibitem [{\citenamefont {Hadad}\ \emph {et~al.}(2017)\citenamefont {Hadad},
  \citenamefont {Vitelli},\ and\ \citenamefont {Alu}}]{Alu2017}%
  \BibitemOpen
  \bibfield  {author} {\bibinfo {author} {\bibfnamefont {Yakir}\ \bibnamefont
  {Hadad}}, \bibinfo {author} {\bibfnamefont {Vincenzo}\ \bibnamefont
  {Vitelli}}, \ and\ \bibinfo {author} {\bibfnamefont {Andrea}\ \bibnamefont
  {Alu}},\ }\bibfield  {title} {\enquote {\bibinfo {title} {Solitons and
  propagating domain walls in topological resonator arrays},}\ }\href {\doibase
  10.1021/acsphotonics.7b00303} {\bibfield  {journal} {\bibinfo  {journal} {ACS
  Photonics}\ }\textbf {\bibinfo {volume} {4}},\ \bibinfo {pages} {1974--1979}
  (\bibinfo {year} {2017})}\BibitemShut {NoStop}%
\bibitem [{\citenamefont {De~Nittis}\ and\ \citenamefont
  {Lein}(2017)}]{Lein2017}%
  \BibitemOpen
  \bibfield  {author} {\bibinfo {author} {\bibfnamefont {Giuseppe}\
  \bibnamefont {De~Nittis}}\ and\ \bibinfo {author} {\bibfnamefont {Max}\
  \bibnamefont {Lein}},\ }\bibfield  {title} {\enquote {\bibinfo {title}
  {Symmetry {Classification} of {Topological} {Photonic} {Crystals}},}\ }\href
  {http://arxiv.org/abs/1710.08104} {\bibfield  {journal} {\bibinfo  {journal}
  {arXiv:1710.08104}\ } (\bibinfo {year} {2017})}\BibitemShut {NoStop}%
\bibitem [{\citenamefont {Lindner}\ \emph {et~al.}(2011)\citenamefont
  {Lindner}, \citenamefont {Refael},\ and\ \citenamefont
  {Galitski}}]{Lindner2011}%
  \BibitemOpen
  \bibfield  {author} {\bibinfo {author} {\bibfnamefont {Netanel~H.}\
  \bibnamefont {Lindner}}, \bibinfo {author} {\bibfnamefont {Gil}\ \bibnamefont
  {Refael}}, \ and\ \bibinfo {author} {\bibfnamefont {Victor}\ \bibnamefont
  {Galitski}},\ }\bibfield  {title} {\enquote {\bibinfo {title} {Floquet
  topological insulator in semiconductor quantum wells},}\ }\href
  {http://dx.doi.org/10.1038/nphys1926} {\bibfield  {journal} {\bibinfo
  {journal} {Nature Physics}\ }\textbf {\bibinfo {volume} {7}},\ \bibinfo
  {pages} {490 EP --} (\bibinfo {year} {2011})},\ \bibinfo {note}
  {article}\BibitemShut {NoStop}%
\bibitem [{\citenamefont {Cayssol}\ \emph {et~al.}(2013)\citenamefont
  {Cayssol}, \citenamefont {Dora}, \citenamefont {Simon},\ and\ \citenamefont
  {Moessner}}]{Cayssol2013}%
  \BibitemOpen
  \bibfield  {author} {\bibinfo {author} {\bibfnamefont {Jerome}\ \bibnamefont
  {Cayssol}}, \bibinfo {author} {\bibfnamefont {Balazs}\ \bibnamefont {Dora}},
  \bibinfo {author} {\bibfnamefont {Ferenc}\ \bibnamefont {Simon}}, \ and\
  \bibinfo {author} {\bibfnamefont {Roderich}\ \bibnamefont {Moessner}},\
  }\bibfield  {title} {\enquote {\bibinfo {title} {Floquet topological
  insulators},}\ }\href {\doibase 10.1002/pssr.201206451} {\bibfield  {journal}
  {\bibinfo  {journal} {physica status solidi (RRL) – Rapid Research
  Letters}\ }\textbf {\bibinfo {volume} {7}},\ \bibinfo {pages} {101--108}
  (\bibinfo {year} {2013})}\BibitemShut {NoStop}%
\bibitem [{\citenamefont {Rechtsman}\ \emph {et~al.}(2013)\citenamefont
  {Rechtsman}, \citenamefont {Zeuner}, \citenamefont {Plotnik}, \citenamefont
  {Lumer}, \citenamefont {Podolsky}, \citenamefont {Dreisow}, \citenamefont
  {Nolte}, \citenamefont {Segev},\ and\ \citenamefont
  {Szameit}}]{Rechtsman2013}%
  \BibitemOpen
  \bibfield  {author} {\bibinfo {author} {\bibfnamefont {Mikael~C.}\
  \bibnamefont {Rechtsman}}, \bibinfo {author} {\bibfnamefont {Julia~M.}\
  \bibnamefont {Zeuner}}, \bibinfo {author} {\bibfnamefont {Yonatan}\
  \bibnamefont {Plotnik}}, \bibinfo {author} {\bibfnamefont {Yaakov}\
  \bibnamefont {Lumer}}, \bibinfo {author} {\bibfnamefont {Daniel}\
  \bibnamefont {Podolsky}}, \bibinfo {author} {\bibfnamefont {Felix}\
  \bibnamefont {Dreisow}}, \bibinfo {author} {\bibfnamefont {Stefan}\
  \bibnamefont {Nolte}}, \bibinfo {author} {\bibfnamefont {Mordechai}\
  \bibnamefont {Segev}}, \ and\ \bibinfo {author} {\bibfnamefont {Alexander}\
  \bibnamefont {Szameit}},\ }\bibfield  {title} {\enquote {\bibinfo {title}
  {Photonic floquet topological insulators},}\ }\href
  {http://dx.doi.org/10.1038/nature12066} {\bibfield  {journal} {\bibinfo
  {journal} {Nature}\ }\textbf {\bibinfo {volume} {496}},\ \bibinfo {pages}
  {196--200} (\bibinfo {year} {2013})},\ \bibinfo {note} {letter}\BibitemShut
  {NoStop}%
\bibitem [{\citenamefont {Khanikaev}\ and\ \citenamefont
  {Shvets}(2017)}]{Khanikaev2017}%
  \BibitemOpen
  \bibfield  {author} {\bibinfo {author} {\bibfnamefont {Alexander~B.}\
  \bibnamefont {Khanikaev}}\ and\ \bibinfo {author} {\bibfnamefont {Gennady}\
  \bibnamefont {Shvets}},\ }\bibfield  {title} {\enquote {\bibinfo {title}
  {Two-dimensional topological photonics},}\ }\href {\doibase
  10.1038/s41566-017-0048-5} {\bibfield  {journal} {\bibinfo  {journal} {Nature
  Photonics}\ }\textbf {\bibinfo {volume} {11}},\ \bibinfo {pages} {763--773}
  (\bibinfo {year} {2017})}\BibitemShut {NoStop}%
\bibitem [{\citenamefont {Slobozhanyuk}\ \emph {et~al.}(2017)\citenamefont
  {Slobozhanyuk}, \citenamefont {Mousavi}, \citenamefont {Ni}, \citenamefont
  {Smirnova}, \citenamefont {Kivshar},\ and\ \citenamefont
  {Khanikaev}}]{Slobozhanyuk2017}%
  \BibitemOpen
  \bibfield  {author} {\bibinfo {author} {\bibfnamefont {Alexey}\ \bibnamefont
  {Slobozhanyuk}}, \bibinfo {author} {\bibfnamefont {S.~Hossein}\ \bibnamefont
  {Mousavi}}, \bibinfo {author} {\bibfnamefont {Xiang}\ \bibnamefont {Ni}},
  \bibinfo {author} {\bibfnamefont {Daria}\ \bibnamefont {Smirnova}}, \bibinfo
  {author} {\bibfnamefont {Yuri~S.}\ \bibnamefont {Kivshar}}, \ and\ \bibinfo
  {author} {\bibfnamefont {Alexander~B.}\ \bibnamefont {Khanikaev}},\
  }\bibfield  {title} {\enquote {\bibinfo {title} {Three-dimensional
  all-dielectric photonic topological insulator},}\ }\href
  {http://dx.doi.org/10.1038/nphoton.2016.253} {\bibfield  {journal} {\bibinfo
  {journal} {Nat Photon}\ }\textbf {\bibinfo {volume} {11}},\ \bibinfo {pages}
  {130--136} (\bibinfo {year} {2017})},\ \bibinfo {note} {article}\BibitemShut
  {NoStop}%
\bibitem [{\citenamefont {He}\ \emph {et~al.}(2016)\citenamefont {He},
  \citenamefont {Sun}, \citenamefont {Liu}, \citenamefont {Lu}, \citenamefont
  {Chen}, \citenamefont {Feng},\ and\ \citenamefont {Chen}}]{He4924}%
  \BibitemOpen
  \bibfield  {author} {\bibinfo {author} {\bibfnamefont {Cheng}\ \bibnamefont
  {He}}, \bibinfo {author} {\bibfnamefont {Xiao-Chen}\ \bibnamefont {Sun}},
  \bibinfo {author} {\bibfnamefont {Xiao-Ping}\ \bibnamefont {Liu}}, \bibinfo
  {author} {\bibfnamefont {Ming-Hui}\ \bibnamefont {Lu}}, \bibinfo {author}
  {\bibfnamefont {Yulin}\ \bibnamefont {Chen}}, \bibinfo {author}
  {\bibfnamefont {Liang}\ \bibnamefont {Feng}}, \ and\ \bibinfo {author}
  {\bibfnamefont {Yan-Feng}\ \bibnamefont {Chen}},\ }\bibfield  {title}
  {\enquote {\bibinfo {title} {Photonic topological insulator with broken
  time-reversal symmetry},}\ }\href {\doibase 10.1073/pnas.1525502113}
  {\bibfield  {journal} {\bibinfo  {journal} {Proceedings of the National
  Academy of Sciences}\ }\textbf {\bibinfo {volume} {113}},\ \bibinfo {pages}
  {4924--4928} (\bibinfo {year} {2016})}\BibitemShut {NoStop}%
\bibitem [{\citenamefont {Glybovski}\ \emph {et~al.}(2016)\citenamefont
  {Glybovski}, \citenamefont {Tretyakov}, \citenamefont {Belov}, \citenamefont
  {Kivshar},\ and\ \citenamefont {Simovski}}]{GLYBOVSKI20161}%
  \BibitemOpen
  \bibfield  {author} {\bibinfo {author} {\bibfnamefont {Stanislav~B.}\
  \bibnamefont {Glybovski}}, \bibinfo {author} {\bibfnamefont {Sergei~A.}\
  \bibnamefont {Tretyakov}}, \bibinfo {author} {\bibfnamefont {Pavel~A.}\
  \bibnamefont {Belov}}, \bibinfo {author} {\bibfnamefont {Yuri~S.}\
  \bibnamefont {Kivshar}}, \ and\ \bibinfo {author} {\bibfnamefont
  {Constantin~R.}\ \bibnamefont {Simovski}},\ }\bibfield  {title} {\enquote
  {\bibinfo {title} {Metasurfaces: From microwaves to visible},}\ }\href
  {\doibase https://doi.org/10.1016/j.physrep.2016.04.004} {\bibfield
  {journal} {\bibinfo  {journal} {Physics Reports}\ }\textbf {\bibinfo {volume}
  {634}},\ \bibinfo {pages} {1 -- 72} (\bibinfo {year} {2016})},\ \bibinfo
  {note} {metasurfaces: From microwaves to visible}\BibitemShut {NoStop}%
\bibitem [{\citenamefont {Lindell}\ and\ \citenamefont
  {Sihvola}(2005)}]{Sihvola2005}%
  \BibitemOpen
  \bibfield  {author} {\bibinfo {author} {\bibfnamefont {I.~V.}\ \bibnamefont
  {Lindell}}\ and\ \bibinfo {author} {\bibfnamefont {A.~H.}\ \bibnamefont
  {Sihvola}},\ }\bibfield  {title} {\enquote {\bibinfo {title} {Realization of
  the pemc boundary},}\ }\href {\doibase 10.1109/TAP.2005.854524} {\bibfield
  {journal} {\bibinfo  {journal} {IEEE Transactions on Antennas and
  Propagation}\ }\textbf {\bibinfo {volume} {53}},\ \bibinfo {pages}
  {3012--3018} (\bibinfo {year} {2005})}\BibitemShut {NoStop}%
\bibitem [{\citenamefont {Alaee}\ \emph {et~al.}(2012)\citenamefont {Alaee},
  \citenamefont {Farhat}, \citenamefont {Rockstuhl},\ and\ \citenamefont
  {Lederer}}]{Alaee:12}%
  \BibitemOpen
  \bibfield  {author} {\bibinfo {author} {\bibfnamefont {Rasoul}\ \bibnamefont
  {Alaee}}, \bibinfo {author} {\bibfnamefont {Mohamed}\ \bibnamefont {Farhat}},
  \bibinfo {author} {\bibfnamefont {Carsten}\ \bibnamefont {Rockstuhl}}, \ and\
  \bibinfo {author} {\bibfnamefont {Falk}\ \bibnamefont {Lederer}},\ }\bibfield
   {title} {\enquote {\bibinfo {title} {A perfect absorber made of a graphene
  micro-ribbon metamaterial},}\ }\href {\doibase 10.1364/OE.20.028017}
  {\bibfield  {journal} {\bibinfo  {journal} {Opt. Express}\ }\textbf {\bibinfo
  {volume} {20}},\ \bibinfo {pages} {28017--28024} (\bibinfo {year}
  {2012})}\BibitemShut {NoStop}%
\bibitem [{\citenamefont {Papasimakis}\ \emph {et~al.}(2010)\citenamefont
  {Papasimakis}, \citenamefont {Luo}, \citenamefont {Shen}, \citenamefont
  {Angelis}, \citenamefont {Fabrizio}, \citenamefont {Nikolaenko},\ and\
  \citenamefont {Zheludev}}]{Papasimakis:10}%
  \BibitemOpen
  \bibfield  {author} {\bibinfo {author} {\bibfnamefont {Nikitas}\ \bibnamefont
  {Papasimakis}}, \bibinfo {author} {\bibfnamefont {Zhiqiang}\ \bibnamefont
  {Luo}}, \bibinfo {author} {\bibfnamefont {Ze~Xiang}\ \bibnamefont {Shen}},
  \bibinfo {author} {\bibfnamefont {Francesco~De}\ \bibnamefont {Angelis}},
  \bibinfo {author} {\bibfnamefont {Enzo~Di}\ \bibnamefont {Fabrizio}},
  \bibinfo {author} {\bibfnamefont {Andrey~E.}\ \bibnamefont {Nikolaenko}}, \
  and\ \bibinfo {author} {\bibfnamefont {Nikolay~I.}\ \bibnamefont
  {Zheludev}},\ }\bibfield  {title} {\enquote {\bibinfo {title} {Graphene in a
  photonic metamaterial},}\ }\href {\doibase 10.1364/OE.18.008353} {\bibfield
  {journal} {\bibinfo  {journal} {Opt. Express}\ }\textbf {\bibinfo {volume}
  {18}},\ \bibinfo {pages} {8353--8359} (\bibinfo {year} {2010})}\BibitemShut
  {NoStop}%
\bibitem [{\citenamefont {Li}\ \emph {et~al.}(2018)\citenamefont {Li},
  \citenamefont {Shen}, \citenamefont {D{\'i}az-Rubio}, \citenamefont
  {Tretyakov},\ and\ \citenamefont {Cummer}}]{Li2018}%
  \BibitemOpen
  \bibfield  {author} {\bibinfo {author} {\bibfnamefont {Junfei}\ \bibnamefont
  {Li}}, \bibinfo {author} {\bibfnamefont {Chen}\ \bibnamefont {Shen}},
  \bibinfo {author} {\bibfnamefont {Ana}\ \bibnamefont {D{\'i}az-Rubio}},
  \bibinfo {author} {\bibfnamefont {Sergei~A.}\ \bibnamefont {Tretyakov}}, \
  and\ \bibinfo {author} {\bibfnamefont {Steven~A.}\ \bibnamefont {Cummer}},\
  }\bibfield  {title} {\enquote {\bibinfo {title} {Systematic design and
  experimental demonstration of bianisotropic metasurfaces for scattering-free
  manipulation of acoustic wavefronts},}\ }\href {\doibase
  10.1038/s41467-018-03778-9} {\bibfield  {journal} {\bibinfo  {journal}
  {Nature Communications}\ }\textbf {\bibinfo {volume} {9}},\ \bibinfo {pages}
  {1342} (\bibinfo {year} {2018})}\BibitemShut {NoStop}%
\bibitem [{\citenamefont {Guo}\ \emph {et~al.}(2017)\citenamefont {Guo},
  \citenamefont {Xiao},\ and\ \citenamefont {Fan}}]{ShanuiFan2017}%
  \BibitemOpen
  \bibfield  {author} {\bibinfo {author} {\bibfnamefont {Yu}~\bibnamefont
  {Guo}}, \bibinfo {author} {\bibfnamefont {Meng}\ \bibnamefont {Xiao}}, \ and\
  \bibinfo {author} {\bibfnamefont {Shanhui}\ \bibnamefont {Fan}},\ }\bibfield
  {title} {\enquote {\bibinfo {title} {Topologically protected complete
  polarization conversion},}\ }\href {\doibase 10.1103/PhysRevLett.119.167401}
  {\bibfield  {journal} {\bibinfo  {journal} {Phys. Rev. Lett.}\ }\textbf
  {\bibinfo {volume} {119}},\ \bibinfo {pages} {167401} (\bibinfo {year}
  {2017})}\BibitemShut {NoStop}%
\bibitem [{\citenamefont {Ding}\ \emph {et~al.}(2016)\citenamefont {Ding},
  \citenamefont {Ma}, \citenamefont {Xiao}, \citenamefont {Zhang},\ and\
  \citenamefont {Chan}}]{CTChan2016}%
  \BibitemOpen
  \bibfield  {author} {\bibinfo {author} {\bibfnamefont {Kun}\ \bibnamefont
  {Ding}}, \bibinfo {author} {\bibfnamefont {Guancong}\ \bibnamefont {Ma}},
  \bibinfo {author} {\bibfnamefont {Meng}\ \bibnamefont {Xiao}}, \bibinfo
  {author} {\bibfnamefont {Z.~Q.}\ \bibnamefont {Zhang}}, \ and\ \bibinfo
  {author} {\bibfnamefont {C.~T.}\ \bibnamefont {Chan}},\ }\bibfield  {title}
  {\enquote {\bibinfo {title} {Emergence, coalescence, and topological
  properties of multiple exceptional points and their experimental
  realization},}\ }\href {\doibase 10.1103/PhysRevX.6.021007} {\bibfield
  {journal} {\bibinfo  {journal} {Phys. Rev. X}\ }\textbf {\bibinfo {volume}
  {6}},\ \bibinfo {pages} {021007} (\bibinfo {year} {2016})}\BibitemShut
  {NoStop}%
\bibitem [{\citenamefont {Silveirinha}(2015)}]{Silveirinha2015}%
  \BibitemOpen
  \bibfield  {author} {\bibinfo {author} {\bibfnamefont {M\'ario~G.}\
  \bibnamefont {Silveirinha}},\ }\bibfield  {title} {\enquote {\bibinfo {title}
  {Chern invariants for continuous media},}\ }\href {\doibase
  10.1103/PhysRevB.92.125153} {\bibfield  {journal} {\bibinfo  {journal} {Phys.
  Rev. B}\ }\textbf {\bibinfo {volume} {92}},\ \bibinfo {pages} {125153}
  (\bibinfo {year} {2015})}\BibitemShut {NoStop}%
\bibitem [{\citenamefont {Gao}\ \emph {et~al.}(2015)\citenamefont {Gao},
  \citenamefont {Lawrence}, \citenamefont {Yang}, \citenamefont {Liu},
  \citenamefont {Fang}, \citenamefont {B\'eri}, \citenamefont {Li},\ and\
  \citenamefont {Zhang}}]{Gao2015}%
  \BibitemOpen
  \bibfield  {author} {\bibinfo {author} {\bibfnamefont {Wenlong}\ \bibnamefont
  {Gao}}, \bibinfo {author} {\bibfnamefont {Mark}\ \bibnamefont {Lawrence}},
  \bibinfo {author} {\bibfnamefont {Biao}\ \bibnamefont {Yang}}, \bibinfo
  {author} {\bibfnamefont {Fu}~\bibnamefont {Liu}}, \bibinfo {author}
  {\bibfnamefont {Fengzhou}\ \bibnamefont {Fang}}, \bibinfo {author}
  {\bibfnamefont {Benjamin}\ \bibnamefont {B\'eri}}, \bibinfo {author}
  {\bibfnamefont {Jensen}\ \bibnamefont {Li}}, \ and\ \bibinfo {author}
  {\bibfnamefont {Shuang}\ \bibnamefont {Zhang}},\ }\bibfield  {title}
  {\enquote {\bibinfo {title} {Topological photonic phase in chiral hyperbolic
  metamaterials},}\ }\href {\doibase 10.1103/PhysRevLett.114.037402} {\bibfield
   {journal} {\bibinfo  {journal} {Phys. Rev. Lett.}\ }\textbf {\bibinfo
  {volume} {114}},\ \bibinfo {pages} {037402} (\bibinfo {year}
  {2015})}\BibitemShut {NoStop}%
\bibitem [{\citenamefont {Jin}\ \emph {et~al.}(2016)\citenamefont {Jin},
  \citenamefont {Lu}, \citenamefont {Wang}, \citenamefont {Fang}, \citenamefont
  {Joannopoulos}, \citenamefont {Soljacic}, \citenamefont {Fu},\ and\
  \citenamefont {Fang}}]{Jin2016}%
  \BibitemOpen
  \bibfield  {author} {\bibinfo {author} {\bibfnamefont {Dafei}\ \bibnamefont
  {Jin}}, \bibinfo {author} {\bibfnamefont {Ling}\ \bibnamefont {Lu}}, \bibinfo
  {author} {\bibfnamefont {Zhong}\ \bibnamefont {Wang}}, \bibinfo {author}
  {\bibfnamefont {Chen}\ \bibnamefont {Fang}}, \bibinfo {author} {\bibfnamefont
  {John~D.}\ \bibnamefont {Joannopoulos}}, \bibinfo {author} {\bibfnamefont
  {Marin}\ \bibnamefont {Soljacic}}, \bibinfo {author} {\bibfnamefont {Liang}\
  \bibnamefont {Fu}}, \ and\ \bibinfo {author} {\bibfnamefont {Nicholas~X.}\
  \bibnamefont {Fang}},\ }\bibfield  {title} {\enquote {\bibinfo {title}
  {Topological magnetoplasmon},}\ }\href
  {http://dx.doi.org/10.1038/ncomms13486} {\bibfield  {journal} {\bibinfo
  {journal} {Nature Communications}\ }\textbf {\bibinfo {volume} {7}},\
  \bibinfo {pages} {13486 EP --} (\bibinfo {year} {2016})},\ \bibinfo {note}
  {article}\BibitemShut {NoStop}%
\bibitem [{\citenamefont {Gangaraj}\ \emph {et~al.}(2017)\citenamefont
  {Gangaraj}, \citenamefont {Silveirinha},\ and\ \citenamefont
  {Hanson}}]{Hassani2017}%
  \BibitemOpen
  \bibfield  {author} {\bibinfo {author} {\bibfnamefont {S.~A.~Hassani}\
  \bibnamefont {Gangaraj}}, \bibinfo {author} {\bibfnamefont {M.~G.}\
  \bibnamefont {Silveirinha}}, \ and\ \bibinfo {author} {\bibfnamefont {G.~W.}\
  \bibnamefont {Hanson}},\ }\bibfield  {title} {\enquote {\bibinfo {title}
  {Berry phase, berry connection, and chern number for a continuum
  bianisotropic material from a classical electromagnetics perspective},}\
  }\href {\doibase 10.1109/JMMCT.2017.2654962} {\bibfield  {journal} {\bibinfo
  {journal} {IEEE Journal on Multiscale and Multiphysics Computational
  Techniques}\ }\textbf {\bibinfo {volume} {2}},\ \bibinfo {pages} {3--17}
  (\bibinfo {year} {2017})}\BibitemShut {NoStop}%
\bibitem [{\citenamefont {Shi}\ and\ \citenamefont {Song}(2018)}]{Song2018}%
  \BibitemOpen
  \bibfield  {author} {\bibinfo {author} {\bibfnamefont {Li-kun}\ \bibnamefont
  {Shi}}\ and\ \bibinfo {author} {\bibfnamefont {Justin C.~W.}\ \bibnamefont
  {Song}},\ }\bibfield  {title} {\enquote {\bibinfo {title} {Plasmon geometric
  phase and plasmon hall shift},}\ }\href {\doibase 10.1103/PhysRevX.8.021020}
  {\bibfield  {journal} {\bibinfo  {journal} {Phys. Rev. X}\ }\textbf {\bibinfo
  {volume} {8}},\ \bibinfo {pages} {021020} (\bibinfo {year}
  {2018})}\BibitemShut {NoStop}%
\bibitem [{\citenamefont {Bialynicki-Birula}\ and\ \citenamefont
  {Bialynicka-Birula}(1987)}]{BB1987}%
  \BibitemOpen
  \bibfield  {author} {\bibinfo {author} {\bibfnamefont {Iwo}\ \bibnamefont
  {Bialynicki-Birula}}\ and\ \bibinfo {author} {\bibfnamefont {Zofia}\
  \bibnamefont {Bialynicka-Birula}},\ }\bibfield  {title} {\enquote {\bibinfo
  {title} {Berry's phase in the relativistic theory of spinning particles},}\
  }\href {\doibase 10.1103/PhysRevD.35.2383} {\bibfield  {journal} {\bibinfo
  {journal} {Phys. Rev. D}\ }\textbf {\bibinfo {volume} {35}},\ \bibinfo
  {pages} {2383--2387} (\bibinfo {year} {1987})}\BibitemShut {NoStop}%
\bibitem [{\citenamefont {Stone}(2016)}]{Stone2016}%
  \BibitemOpen
  \bibfield  {author} {\bibinfo {author} {\bibfnamefont {Michael}\ \bibnamefont
  {Stone}},\ }\bibfield  {title} {\enquote {\bibinfo {title} {Berry phase and
  anomalous velocity of weyl fermions and maxwell photons},}\ }\href {\doibase
  10.1142/S0217979215502495} {\bibfield  {journal} {\bibinfo  {journal}
  {International Journal of Modern Physics B}\ }\textbf {\bibinfo {volume}
  {30}},\ \bibinfo {pages} {1550249} (\bibinfo {year} {2016})}\BibitemShut
  {NoStop}%
\bibitem [{\citenamefont {Stone}(2015)}]{Stone1432}%
  \BibitemOpen
  \bibfield  {author} {\bibinfo {author} {\bibfnamefont {Michael}\ \bibnamefont
  {Stone}},\ }\bibfield  {title} {\enquote {\bibinfo {title} {Topology, spin,
  and light},}\ }\href {\doibase 10.1126/science.aac4368} {\bibfield  {journal}
  {\bibinfo  {journal} {Science}\ }\textbf {\bibinfo {volume} {348}},\ \bibinfo
  {pages} {1432--1433} (\bibinfo {year} {2015})}\BibitemShut {NoStop}%
\bibitem [{\citenamefont {Gawhary}\ \emph {et~al.}(2018)\citenamefont
  {Gawhary}, \citenamefont {Van~Mechelen},\ and\ \citenamefont
  {Urbach}}]{Gawhary2018}%
  \BibitemOpen
  \bibfield  {author} {\bibinfo {author} {\bibfnamefont {O.~El}\ \bibnamefont
  {Gawhary}}, \bibinfo {author} {\bibfnamefont {T.}~\bibnamefont
  {Van~Mechelen}}, \ and\ \bibinfo {author} {\bibfnamefont {H.~P.}\
  \bibnamefont {Urbach}},\ }\bibfield  {title} {\enquote {\bibinfo {title}
  {Role of radial charges on the angular momentum of electromagnetic fields:
  Spin-$3/2$ light},}\ }\href {\doibase 10.1103/PhysRevLett.121.123202}
  {\bibfield  {journal} {\bibinfo  {journal} {Phys. Rev. Lett.}\ }\textbf
  {\bibinfo {volume} {121}},\ \bibinfo {pages} {123202} (\bibinfo {year}
  {2018})}\BibitemShut {NoStop}%
\bibitem [{\citenamefont {Hubbard}(1955)}]{Hubbard1955}%
  \BibitemOpen
  \bibfield  {author} {\bibinfo {author} {\bibfnamefont {J}~\bibnamefont
  {Hubbard}},\ }\bibfield  {title} {\enquote {\bibinfo {title} {The dielectric
  theory of electronic interactions in solids},}\ }\href
  {http://stacks.iop.org/0370-1298/68/i=11/a=304} {\bibfield  {journal}
  {\bibinfo  {journal} {Proceedings of the Physical Society. Section A}\
  }\textbf {\bibinfo {volume} {68}},\ \bibinfo {pages} {976} (\bibinfo {year}
  {1955})}\BibitemShut {NoStop}%
\bibitem [{\citenamefont {Falk}(1960)}]{Falk1960}%
  \BibitemOpen
  \bibfield  {author} {\bibinfo {author} {\bibfnamefont {David~S.}\
  \bibnamefont {Falk}},\ }\bibfield  {title} {\enquote {\bibinfo {title}
  {Effect of the lattice on dielectric properties of an electron gas},}\ }\href
  {\doibase 10.1103/PhysRev.118.105} {\bibfield  {journal} {\bibinfo  {journal}
  {Phys. Rev.}\ }\textbf {\bibinfo {volume} {118}},\ \bibinfo {pages}
  {105--109} (\bibinfo {year} {1960})}\BibitemShut {NoStop}%
\bibitem [{\citenamefont {Penn}(1962)}]{Penn1962}%
  \BibitemOpen
  \bibfield  {author} {\bibinfo {author} {\bibfnamefont {David~R.}\
  \bibnamefont {Penn}},\ }\bibfield  {title} {\enquote {\bibinfo {title}
  {Wave-number-dependent dielectric function of semiconductors},}\ }\href
  {\doibase 10.1103/PhysRev.128.2093} {\bibfield  {journal} {\bibinfo
  {journal} {Phys. Rev.}\ }\textbf {\bibinfo {volume} {128}},\ \bibinfo {pages}
  {2093--2097} (\bibinfo {year} {1962})}\BibitemShut {NoStop}%
\bibitem [{\citenamefont {Landau}\ \emph {et~al.}(2013)\citenamefont {Landau},
  \citenamefont {Bell}, \citenamefont {Kearsley}, \citenamefont {Pitaevskii},
  \citenamefont {Lifshitz},\ and\ \citenamefont
  {Sykes}}]{landau2013electrodynamics}%
  \BibitemOpen
  \bibfield  {author} {\bibinfo {author} {\bibfnamefont {Lev~Davidovich}\
  \bibnamefont {Landau}}, \bibinfo {author} {\bibfnamefont {JS}~\bibnamefont
  {Bell}}, \bibinfo {author} {\bibfnamefont {MJ}~\bibnamefont {Kearsley}},
  \bibinfo {author} {\bibfnamefont {LP}~\bibnamefont {Pitaevskii}}, \bibinfo
  {author} {\bibfnamefont {EM}~\bibnamefont {Lifshitz}}, \ and\ \bibinfo
  {author} {\bibfnamefont {JB}~\bibnamefont {Sykes}},\ }\href@noop {} {\emph
  {\bibinfo {title} {Electrodynamics of Continuous Media}}},\ Vol.~\bibinfo
  {volume} {8}\ (\bibinfo  {publisher} {Elsevier},\ \bibinfo {address} {New
  York},\ \bibinfo {year} {2013})\BibitemShut {NoStop}%
\bibitem [{\citenamefont {Agranovich}\ and\ \citenamefont
  {Ginzburg}(2013)}]{agranovich2013crystal}%
  \BibitemOpen
  \bibfield  {author} {\bibinfo {author} {\bibfnamefont {Vladimir~M}\
  \bibnamefont {Agranovich}}\ and\ \bibinfo {author} {\bibfnamefont {Vitaly}\
  \bibnamefont {Ginzburg}},\ }\href@noop {} {\emph {\bibinfo {title} {Crystal
  optics with spatial dispersion, and excitons}}},\ Vol.~\bibinfo {volume}
  {42}\ (\bibinfo  {publisher} {Springer Science \& Business Media, New York},\
  \bibinfo {year} {2013})\BibitemShut {NoStop}%
\bibitem [{\citenamefont {Fu}(2011)}]{Fu_Liang2011}%
  \BibitemOpen
  \bibfield  {author} {\bibinfo {author} {\bibfnamefont {Liang}\ \bibnamefont
  {Fu}},\ }\bibfield  {title} {\enquote {\bibinfo {title} {Topological
  crystalline insulators},}\ }\href {\doibase 10.1103/PhysRevLett.106.106802}
  {\bibfield  {journal} {\bibinfo  {journal} {Phys. Rev. Lett.}\ }\textbf
  {\bibinfo {volume} {106}},\ \bibinfo {pages} {106802} (\bibinfo {year}
  {2011})}\BibitemShut {NoStop}%
\bibitem [{\citenamefont {Hughes}\ \emph {et~al.}(2011)\citenamefont {Hughes},
  \citenamefont {Prodan},\ and\ \citenamefont {Bernevig}}]{Hughes2011}%
  \BibitemOpen
  \bibfield  {author} {\bibinfo {author} {\bibfnamefont {Taylor~L.}\
  \bibnamefont {Hughes}}, \bibinfo {author} {\bibfnamefont {Emil}\ \bibnamefont
  {Prodan}}, \ and\ \bibinfo {author} {\bibfnamefont {B.~Andrei}\ \bibnamefont
  {Bernevig}},\ }\bibfield  {title} {\enquote {\bibinfo {title}
  {Inversion-symmetric topological insulators},}\ }\href {\doibase
  10.1103/PhysRevB.83.245132} {\bibfield  {journal} {\bibinfo  {journal} {Phys.
  Rev. B}\ }\textbf {\bibinfo {volume} {83}},\ \bibinfo {pages} {245132}
  (\bibinfo {year} {2011})}\BibitemShut {NoStop}%
\bibitem [{\citenamefont {Slager}\ \emph {et~al.}(2012)\citenamefont {Slager},
  \citenamefont {Mesaros}, \citenamefont {Juricic},\ and\ \citenamefont
  {Zaanen}}]{Slager2012}%
  \BibitemOpen
  \bibfield  {author} {\bibinfo {author} {\bibfnamefont {Robert-Jan}\
  \bibnamefont {Slager}}, \bibinfo {author} {\bibfnamefont {Andrej}\
  \bibnamefont {Mesaros}}, \bibinfo {author} {\bibfnamefont {Vladimir}\
  \bibnamefont {Juricic}}, \ and\ \bibinfo {author} {\bibfnamefont {Jan}\
  \bibnamefont {Zaanen}},\ }\bibfield  {title} {\enquote {\bibinfo {title} {The
  space group classification of topological band-insulators},}\ }\href
  {http://dx.doi.org/10.1038/nphys2513} {\bibfield  {journal} {\bibinfo
  {journal} {Nature Physics}\ }\textbf {\bibinfo {volume} {9}},\ \bibinfo
  {pages} {98 EP --} (\bibinfo {year} {2012})},\ \bibinfo {note}
  {article}\BibitemShut {NoStop}%
\bibitem [{\citenamefont {Fang}\ \emph {et~al.}(2012)\citenamefont {Fang},
  \citenamefont {Gilbert},\ and\ \citenamefont {Bernevig}}]{Fang2012}%
  \BibitemOpen
  \bibfield  {author} {\bibinfo {author} {\bibfnamefont {Chen}\ \bibnamefont
  {Fang}}, \bibinfo {author} {\bibfnamefont {Matthew~J.}\ \bibnamefont
  {Gilbert}}, \ and\ \bibinfo {author} {\bibfnamefont {B.~Andrei}\ \bibnamefont
  {Bernevig}},\ }\bibfield  {title} {\enquote {\bibinfo {title} {Bulk
  topological invariants in noninteracting point group symmetric insulators},}\
  }\href {\doibase 10.1103/PhysRevB.86.115112} {\bibfield  {journal} {\bibinfo
  {journal} {Phys. Rev. B}\ }\textbf {\bibinfo {volume} {86}},\ \bibinfo
  {pages} {115112} (\bibinfo {year} {2012})}\BibitemShut {NoStop}%
\bibitem [{\citenamefont {Kruthoff}\ \emph {et~al.}(2017)\citenamefont
  {Kruthoff}, \citenamefont {de~Boer}, \citenamefont {van Wezel}, \citenamefont
  {Kane},\ and\ \citenamefont {Slager}}]{Kruthoff2017}%
  \BibitemOpen
  \bibfield  {author} {\bibinfo {author} {\bibfnamefont {Jorrit}\ \bibnamefont
  {Kruthoff}}, \bibinfo {author} {\bibfnamefont {Jan}\ \bibnamefont {de~Boer}},
  \bibinfo {author} {\bibfnamefont {Jasper}\ \bibnamefont {van Wezel}},
  \bibinfo {author} {\bibfnamefont {Charles~L.}\ \bibnamefont {Kane}}, \ and\
  \bibinfo {author} {\bibfnamefont {Robert-Jan}\ \bibnamefont {Slager}},\
  }\bibfield  {title} {\enquote {\bibinfo {title} {Topological classification
  of crystalline insulators through band structure combinatorics},}\ }\href
  {\doibase 10.1103/PhysRevX.7.041069} {\bibfield  {journal} {\bibinfo
  {journal} {Phys. Rev. X}\ }\textbf {\bibinfo {volume} {7}},\ \bibinfo {pages}
  {041069} (\bibinfo {year} {2017})}\BibitemShut {NoStop}%
\bibitem [{\citenamefont {Bradlyn}\ \emph {et~al.}(2017)\citenamefont
  {Bradlyn}, \citenamefont {Elcoro}, \citenamefont {Cano}, \citenamefont
  {Vergniory}, \citenamefont {Wang}, \citenamefont {Felser}, \citenamefont
  {Aroyo},\ and\ \citenamefont {Bernevig}}]{Bradlyn2017}%
  \BibitemOpen
  \bibfield  {author} {\bibinfo {author} {\bibfnamefont {Barry}\ \bibnamefont
  {Bradlyn}}, \bibinfo {author} {\bibfnamefont {L.}~\bibnamefont {Elcoro}},
  \bibinfo {author} {\bibfnamefont {Jennifer}\ \bibnamefont {Cano}}, \bibinfo
  {author} {\bibfnamefont {M.~G.}\ \bibnamefont {Vergniory}}, \bibinfo {author}
  {\bibfnamefont {Zhijun}\ \bibnamefont {Wang}}, \bibinfo {author}
  {\bibfnamefont {C.}~\bibnamefont {Felser}}, \bibinfo {author} {\bibfnamefont
  {M.~I.}\ \bibnamefont {Aroyo}}, \ and\ \bibinfo {author} {\bibfnamefont
  {B.~Andrei}\ \bibnamefont {Bernevig}},\ }\bibfield  {title} {\enquote
  {\bibinfo {title} {Topological quantum chemistry},}\ }\href
  {http://dx.doi.org/10.1038/nature23268} {\bibfield  {journal} {\bibinfo
  {journal} {Nature}\ }\textbf {\bibinfo {volume} {547}},\ \bibinfo {pages}
  {298 EP --} (\bibinfo {year} {2017})},\ \bibinfo {note} {article}\BibitemShut
  {NoStop}%
\bibitem [{\citenamefont {Song}\ \emph {et~al.}(2017)\citenamefont {Song},
  \citenamefont {Huang}, \citenamefont {Fu},\ and\ \citenamefont
  {Hermele}}]{Song2017}%
  \BibitemOpen
  \bibfield  {author} {\bibinfo {author} {\bibfnamefont {Hao}\ \bibnamefont
  {Song}}, \bibinfo {author} {\bibfnamefont {Sheng-Jie}\ \bibnamefont {Huang}},
  \bibinfo {author} {\bibfnamefont {Liang}\ \bibnamefont {Fu}}, \ and\ \bibinfo
  {author} {\bibfnamefont {Michael}\ \bibnamefont {Hermele}},\ }\bibfield
  {title} {\enquote {\bibinfo {title} {Topological phases protected by point
  group symmetry},}\ }\href {\doibase 10.1103/PhysRevX.7.011020} {\bibfield
  {journal} {\bibinfo  {journal} {Phys. Rev. X}\ }\textbf {\bibinfo {volume}
  {7}},\ \bibinfo {pages} {011020} (\bibinfo {year} {2017})}\BibitemShut
  {NoStop}%
\bibitem [{\citenamefont {Sedrakyan}\ \emph {et~al.}(2017)\citenamefont
  {Sedrakyan}, \citenamefont {Galitski},\ and\ \citenamefont
  {Kamenev}}]{Sedrakyan2017}%
  \BibitemOpen
  \bibfield  {author} {\bibinfo {author} {\bibfnamefont {Tigran~A.}\
  \bibnamefont {Sedrakyan}}, \bibinfo {author} {\bibfnamefont {Victor~M.}\
  \bibnamefont {Galitski}}, \ and\ \bibinfo {author} {\bibfnamefont {Alex}\
  \bibnamefont {Kamenev}},\ }\bibfield  {title} {\enquote {\bibinfo {title}
  {Topological spin ordering via chern-simons superconductivity},}\ }\href
  {\doibase 10.1103/PhysRevB.95.094511} {\bibfield  {journal} {\bibinfo
  {journal} {Phys. Rev. B}\ }\textbf {\bibinfo {volume} {95}},\ \bibinfo
  {pages} {094511} (\bibinfo {year} {2017})}\BibitemShut {NoStop}%
\bibitem [{\citenamefont {Po}\ \emph {et~al.}(2017)\citenamefont {Po},
  \citenamefont {Vishwanath},\ and\ \citenamefont {Watanabe}}]{Po2017}%
  \BibitemOpen
  \bibfield  {author} {\bibinfo {author} {\bibfnamefont {Hoi~Chun}\
  \bibnamefont {Po}}, \bibinfo {author} {\bibfnamefont {Ashvin}\ \bibnamefont
  {Vishwanath}}, \ and\ \bibinfo {author} {\bibfnamefont {Haruki}\ \bibnamefont
  {Watanabe}},\ }\bibfield  {title} {\enquote {\bibinfo {title} {Symmetry-based
  indicators of band topology in the 230 space groups},}\ }\href {\doibase
  10.1038/s41467-017-00133-2} {\bibfield  {journal} {\bibinfo  {journal}
  {Nature Communications}\ }\textbf {\bibinfo {volume} {8}},\ \bibinfo {pages}
  {50} (\bibinfo {year} {2017})}\BibitemShut {NoStop}%
\bibitem [{\citenamefont {Thorngren}\ and\ \citenamefont
  {Else}(2018)}]{Thorngren2018}%
  \BibitemOpen
  \bibfield  {author} {\bibinfo {author} {\bibfnamefont {Ryan}\ \bibnamefont
  {Thorngren}}\ and\ \bibinfo {author} {\bibfnamefont {Dominic~V.}\
  \bibnamefont {Else}},\ }\bibfield  {title} {\enquote {\bibinfo {title}
  {Gauging spatial symmetries and the classification of topological crystalline
  phases},}\ }\href {\doibase 10.1103/PhysRevX.8.011040} {\bibfield  {journal}
  {\bibinfo  {journal} {Phys. Rev. X}\ }\textbf {\bibinfo {volume} {8}},\
  \bibinfo {pages} {011040} (\bibinfo {year} {2018})}\BibitemShut {NoStop}%
\bibitem [{\citenamefont {Matsugatani}\ \emph {et~al.}(2018)\citenamefont
  {Matsugatani}, \citenamefont {Ishiguro}, \citenamefont {Shiozaki},\ and\
  \citenamefont {Watanabe}}]{Matsugatani2018}%
  \BibitemOpen
  \bibfield  {author} {\bibinfo {author} {\bibfnamefont {Akishi}\ \bibnamefont
  {Matsugatani}}, \bibinfo {author} {\bibfnamefont {Yuri}\ \bibnamefont
  {Ishiguro}}, \bibinfo {author} {\bibfnamefont {Ken}\ \bibnamefont
  {Shiozaki}}, \ and\ \bibinfo {author} {\bibfnamefont {Haruki}\ \bibnamefont
  {Watanabe}},\ }\bibfield  {title} {\enquote {\bibinfo {title} {Universal
  relation among the many-body chern number, rotation symmetry, and filling},}\
  }\href {\doibase 10.1103/PhysRevLett.120.096601} {\bibfield  {journal}
  {\bibinfo  {journal} {Phys. Rev. Lett.}\ }\textbf {\bibinfo {volume} {120}},\
  \bibinfo {pages} {096601} (\bibinfo {year} {2018})}\BibitemShut {NoStop}%
\bibitem [{\citenamefont {Watanabe}\ \emph {et~al.}(2018)\citenamefont
  {Watanabe}, \citenamefont {Po},\ and\ \citenamefont
  {Vishwanath}}]{Watanabet2018}%
  \BibitemOpen
  \bibfield  {author} {\bibinfo {author} {\bibfnamefont {Haruki}\ \bibnamefont
  {Watanabe}}, \bibinfo {author} {\bibfnamefont {Hoi~Chun}\ \bibnamefont {Po}},
  \ and\ \bibinfo {author} {\bibfnamefont {Ashvin}\ \bibnamefont
  {Vishwanath}},\ }\bibfield  {title} {\enquote {\bibinfo {title} {Structure
  and topology of band structures in the 1651 magnetic space groups},}\ }\href
  {http://advances.sciencemag.org/content/4/8/eaat8685} {\bibfield  {journal}
  {\bibinfo  {journal} {Science Advances}\ }\textbf {\bibinfo {volume} {4}}
  (\bibinfo {year} {2018})}\BibitemShut {NoStop}%
\bibitem [{\citenamefont {Bradlyn}\ \emph {et~al.}(2016)\citenamefont
  {Bradlyn}, \citenamefont {Cano}, \citenamefont {Wang}, \citenamefont
  {Vergniory}, \citenamefont {Felser}, \citenamefont {Cava},\ and\
  \citenamefont {Bernevig}}]{Bradlynaaf5037}%
  \BibitemOpen
  \bibfield  {author} {\bibinfo {author} {\bibfnamefont {Barry}\ \bibnamefont
  {Bradlyn}}, \bibinfo {author} {\bibfnamefont {Jennifer}\ \bibnamefont
  {Cano}}, \bibinfo {author} {\bibfnamefont {Zhijun}\ \bibnamefont {Wang}},
  \bibinfo {author} {\bibfnamefont {M.~G.}\ \bibnamefont {Vergniory}}, \bibinfo
  {author} {\bibfnamefont {C.}~\bibnamefont {Felser}}, \bibinfo {author}
  {\bibfnamefont {R.~J.}\ \bibnamefont {Cava}}, \ and\ \bibinfo {author}
  {\bibfnamefont {B.~Andrei}\ \bibnamefont {Bernevig}},\ }\bibfield  {title}
  {\enquote {\bibinfo {title} {Beyond dirac and weyl fermions: Unconventional
  quasiparticles in conventional crystals},}\ }\href
  {http://science.sciencemag.org/content/353/6299/aaf5037} {\bibfield
  {journal} {\bibinfo  {journal} {Science}\ }\textbf {\bibinfo {volume} {353}}
  (\bibinfo {year} {2016})}\BibitemShut {NoStop}%
\bibitem [{\citenamefont {Zhu}\ \emph {et~al.}(2017)\citenamefont {Zhu},
  \citenamefont {Zhang}, \citenamefont {Yan}, \citenamefont {Xing},\ and\
  \citenamefont {Zhu}}]{Zhu2017}%
  \BibitemOpen
  \bibfield  {author} {\bibinfo {author} {\bibfnamefont {Yan-Qing}\
  \bibnamefont {Zhu}}, \bibinfo {author} {\bibfnamefont {Dan-Wei}\ \bibnamefont
  {Zhang}}, \bibinfo {author} {\bibfnamefont {Hui}\ \bibnamefont {Yan}},
  \bibinfo {author} {\bibfnamefont {Ding-Yu}\ \bibnamefont {Xing}}, \ and\
  \bibinfo {author} {\bibfnamefont {Shi-Liang}\ \bibnamefont {Zhu}},\
  }\bibfield  {title} {\enquote {\bibinfo {title} {Emergent pseudospin-1
  maxwell fermions with a threefold degeneracy in optical lattices},}\ }\href
  {\doibase 10.1103/PhysRevA.96.033634} {\bibfield  {journal} {\bibinfo
  {journal} {Phys. Rev. A}\ }\textbf {\bibinfo {volume} {96}},\ \bibinfo
  {pages} {033634} (\bibinfo {year} {2017})}\BibitemShut {NoStop}%
\bibitem [{\citenamefont {Hu}\ and\ \citenamefont {Zhang}(2018)}]{Hu2018}%
  \BibitemOpen
  \bibfield  {author} {\bibinfo {author} {\bibfnamefont {Haiping}\ \bibnamefont
  {Hu}}\ and\ \bibinfo {author} {\bibfnamefont {Chuanwei}\ \bibnamefont
  {Zhang}},\ }\bibfield  {title} {\enquote {\bibinfo {title} {Spin-1
  topological monopoles in the parameter space of ultracold atoms},}\ }\href
  {\doibase 10.1103/PhysRevA.98.013627} {\bibfield  {journal} {\bibinfo
  {journal} {Phys. Rev. A}\ }\textbf {\bibinfo {volume} {98}},\ \bibinfo
  {pages} {013627} (\bibinfo {year} {2018})}\BibitemShut {NoStop}%
\bibitem [{\citenamefont {Hu}\ \emph {et~al.}(2018)\citenamefont {Hu},
  \citenamefont {Hou}, \citenamefont {Zhang},\ and\ \citenamefont
  {Zhang}}]{Hu2018_2}%
  \BibitemOpen
  \bibfield  {author} {\bibinfo {author} {\bibfnamefont {Haiping}\ \bibnamefont
  {Hu}}, \bibinfo {author} {\bibfnamefont {Junpeng}\ \bibnamefont {Hou}},
  \bibinfo {author} {\bibfnamefont {Fan}\ \bibnamefont {Zhang}}, \ and\
  \bibinfo {author} {\bibfnamefont {Chuanwei}\ \bibnamefont {Zhang}},\
  }\bibfield  {title} {\enquote {\bibinfo {title} {Topological triply
  degenerate points induced by spin-tensor-momentum couplings},}\ }\href
  {\doibase 10.1103/PhysRevLett.120.240401} {\bibfield  {journal} {\bibinfo
  {journal} {Phys. Rev. Lett.}\ }\textbf {\bibinfo {volume} {120}},\ \bibinfo
  {pages} {240401} (\bibinfo {year} {2018})}\BibitemShut {NoStop}%
\bibitem [{\citenamefont {Fulga}\ \emph {et~al.}(2018)\citenamefont {Fulga},
  \citenamefont {Fallani},\ and\ \citenamefont {Burrello}}]{Fulga2018}%
  \BibitemOpen
  \bibfield  {author} {\bibinfo {author} {\bibfnamefont {I.~C.}\ \bibnamefont
  {Fulga}}, \bibinfo {author} {\bibfnamefont {L.}~\bibnamefont {Fallani}}, \
  and\ \bibinfo {author} {\bibfnamefont {M.}~\bibnamefont {Burrello}},\
  }\bibfield  {title} {\enquote {\bibinfo {title} {Geometrically protected
  triple-point crossings in an optical lattice},}\ }\href {\doibase
  10.1103/PhysRevB.97.121402} {\bibfield  {journal} {\bibinfo  {journal} {Phys.
  Rev. B}\ }\textbf {\bibinfo {volume} {97}},\ \bibinfo {pages} {121402}
  (\bibinfo {year} {2018})}\BibitemShut {NoStop}%
\bibitem [{\citenamefont {Tan}\ \emph {et~al.}(2018)\citenamefont {Tan},
  \citenamefont {Zhang}, \citenamefont {Liu}, \citenamefont {Xue},
  \citenamefont {Yu}, \citenamefont {Zhu}, \citenamefont {Yan}, \citenamefont
  {Zhu},\ and\ \citenamefont {Yu}}]{Tan2018}%
  \BibitemOpen
  \bibfield  {author} {\bibinfo {author} {\bibfnamefont {Xinsheng}\
  \bibnamefont {Tan}}, \bibinfo {author} {\bibfnamefont {Dan-Wei}\ \bibnamefont
  {Zhang}}, \bibinfo {author} {\bibfnamefont {Qiang}\ \bibnamefont {Liu}},
  \bibinfo {author} {\bibfnamefont {Guangming}\ \bibnamefont {Xue}}, \bibinfo
  {author} {\bibfnamefont {Hai-Feng}\ \bibnamefont {Yu}}, \bibinfo {author}
  {\bibfnamefont {Yan-Qing}\ \bibnamefont {Zhu}}, \bibinfo {author}
  {\bibfnamefont {Hui}\ \bibnamefont {Yan}}, \bibinfo {author} {\bibfnamefont
  {Shi-Liang}\ \bibnamefont {Zhu}}, \ and\ \bibinfo {author} {\bibfnamefont
  {Yang}\ \bibnamefont {Yu}},\ }\bibfield  {title} {\enquote {\bibinfo {title}
  {Topological maxwell metal bands in a superconducting qutrit},}\ }\href
  {\doibase 10.1103/PhysRevLett.120.130503} {\bibfield  {journal} {\bibinfo
  {journal} {Phys. Rev. Lett.}\ }\textbf {\bibinfo {volume} {120}},\ \bibinfo
  {pages} {130503} (\bibinfo {year} {2018})}\BibitemShut {NoStop}%
\bibitem [{\citenamefont {Van~Mechelen}\ and\ \citenamefont
  {Jacob}(2017)}]{van_mechelen2017}%
  \BibitemOpen
  \bibfield  {author} {\bibinfo {author} {\bibfnamefont {Todd}\ \bibnamefont
  {Van~Mechelen}}\ and\ \bibinfo {author} {\bibfnamefont {Zubin}\ \bibnamefont
  {Jacob}},\ }\bibfield  {title} {\enquote {\bibinfo {title} {Dirac-{Maxwell}
  correspondence: {Spin}-1 bosonic topological insulator},}\ }\href
  {http://arxiv.org/abs/1708.08192} {\bibfield  {journal} {\bibinfo  {journal}
  {arXiv: 1708.08192}\ } (\bibinfo {year} {2017})}\BibitemShut {NoStop}%
\bibitem [{\citenamefont {Yin}\ \emph {et~al.}(2013)\citenamefont {Yin},
  \citenamefont {Ye}, \citenamefont {Rho}, \citenamefont {Wang},\ and\
  \citenamefont {Zhang}}]{Yin1405}%
  \BibitemOpen
  \bibfield  {author} {\bibinfo {author} {\bibfnamefont {Xiaobo}\ \bibnamefont
  {Yin}}, \bibinfo {author} {\bibfnamefont {Ziliang}\ \bibnamefont {Ye}},
  \bibinfo {author} {\bibfnamefont {Junsuk}\ \bibnamefont {Rho}}, \bibinfo
  {author} {\bibfnamefont {Yuan}\ \bibnamefont {Wang}}, \ and\ \bibinfo
  {author} {\bibfnamefont {Xiang}\ \bibnamefont {Zhang}},\ }\bibfield  {title}
  {\enquote {\bibinfo {title} {Photonic spin hall effect at metasurfaces},}\
  }\href {\doibase 10.1126/science.1231758} {\bibfield  {journal} {\bibinfo
  {journal} {Science}\ }\textbf {\bibinfo {volume} {339}},\ \bibinfo {pages}
  {1405--1407} (\bibinfo {year} {2013})}\BibitemShut {NoStop}%
\bibitem [{\citenamefont {Mechelen}\ and\ \citenamefont
  {Jacob}(2016)}]{VanMechelen:16}%
  \BibitemOpen
  \bibfield  {author} {\bibinfo {author} {\bibfnamefont {Todd~Van}\
  \bibnamefont {Mechelen}}\ and\ \bibinfo {author} {\bibfnamefont {Zubin}\
  \bibnamefont {Jacob}},\ }\bibfield  {title} {\enquote {\bibinfo {title}
  {Universal spin-momentum locking of evanescent waves},}\ }\href {\doibase
  10.1364/OPTICA.3.000118} {\bibfield  {journal} {\bibinfo  {journal} {Optica}\
  }\textbf {\bibinfo {volume} {3}},\ \bibinfo {pages} {118--126} (\bibinfo
  {year} {2016})}\BibitemShut {NoStop}%
\bibitem [{\citenamefont {Kalhor}\ \emph {et~al.}(2016)\citenamefont {Kalhor},
  \citenamefont {Thundat},\ and\ \citenamefont
  {Jacob}}]{kalhor_universal_2016}%
  \BibitemOpen
  \bibfield  {author} {\bibinfo {author} {\bibfnamefont {Farid}\ \bibnamefont
  {Kalhor}}, \bibinfo {author} {\bibfnamefont {Thomas}\ \bibnamefont
  {Thundat}}, \ and\ \bibinfo {author} {\bibfnamefont {Zubin}\ \bibnamefont
  {Jacob}},\ }\bibfield  {title} {\enquote {\bibinfo {title} {Universal
  spin-momentum locked optical forces},}\ }\href {\doibase 10.1063/1.4941539}
  {\bibfield  {journal} {\bibinfo  {journal} {Applied Physics Letters}\
  }\textbf {\bibinfo {volume} {108}},\ \bibinfo {pages} {061102} (\bibinfo
  {year} {2016})}\BibitemShut {NoStop}%
\bibitem [{\citenamefont {Pendharker}\ \emph {et~al.}(2018)\citenamefont
  {Pendharker}, \citenamefont {Kalhor}, \citenamefont {Mechelen}, \citenamefont
  {Jahani}, \citenamefont {Nazemifard}, \citenamefont {Thundat},\ and\
  \citenamefont {Jacob}}]{Pendharker:18}%
  \BibitemOpen
  \bibfield  {author} {\bibinfo {author} {\bibfnamefont {Sarang}\ \bibnamefont
  {Pendharker}}, \bibinfo {author} {\bibfnamefont {Farid}\ \bibnamefont
  {Kalhor}}, \bibinfo {author} {\bibfnamefont {Todd~Van}\ \bibnamefont
  {Mechelen}}, \bibinfo {author} {\bibfnamefont {Saman}\ \bibnamefont
  {Jahani}}, \bibinfo {author} {\bibfnamefont {Neda}\ \bibnamefont
  {Nazemifard}}, \bibinfo {author} {\bibfnamefont {Thomas}\ \bibnamefont
  {Thundat}}, \ and\ \bibinfo {author} {\bibfnamefont {Zubin}\ \bibnamefont
  {Jacob}},\ }\bibfield  {title} {\enquote {\bibinfo {title} {Spin photonic
  forces in non-reciprocal waveguides},}\ }\href {\doibase
  10.1364/OE.26.023898} {\bibfield  {journal} {\bibinfo  {journal} {Opt.
  Express}\ }\textbf {\bibinfo {volume} {26}},\ \bibinfo {pages} {23898--23910}
  (\bibinfo {year} {2018})}\BibitemShut {NoStop}%
\bibitem [{\citenamefont {Bliokh}\ \emph {et~al.}(2015)\citenamefont {Bliokh},
  \citenamefont {Smirnova},\ and\ \citenamefont {Nori}}]{Bliokh1448}%
  \BibitemOpen
  \bibfield  {author} {\bibinfo {author} {\bibfnamefont {Konstantin~Y.}\
  \bibnamefont {Bliokh}}, \bibinfo {author} {\bibfnamefont {Daria}\
  \bibnamefont {Smirnova}}, \ and\ \bibinfo {author} {\bibfnamefont {Franco}\
  \bibnamefont {Nori}},\ }\bibfield  {title} {\enquote {\bibinfo {title}
  {Quantum spin hall effect of light},}\ }\href {\doibase
  10.1126/science.aaa9519} {\bibfield  {journal} {\bibinfo  {journal}
  {Science}\ }\textbf {\bibinfo {volume} {348}},\ \bibinfo {pages} {1448--1451}
  (\bibinfo {year} {2015})}\BibitemShut {NoStop}%
\bibitem [{\citenamefont {Born}\ and\ \citenamefont
  {Huang}(1954)}]{born1954dynamical}%
  \BibitemOpen
  \bibfield  {author} {\bibinfo {author} {\bibfnamefont {Max}\ \bibnamefont
  {Born}}\ and\ \bibinfo {author} {\bibfnamefont {Kun}\ \bibnamefont {Huang}},\
  }\href@noop {} {\emph {\bibinfo {title} {Dynamical theory of crystal
  lattices}}}\ (\bibinfo  {publisher} {Clarendon Press, Oxford},\ \bibinfo
  {year} {1954})\BibitemShut {NoStop}%
\bibitem [{\citenamefont {Horsley}\ and\ \citenamefont
  {Philbin}(2014)}]{Philbin2014}%
  \BibitemOpen
  \bibfield  {author} {\bibinfo {author} {\bibfnamefont {S~A~R}\ \bibnamefont
  {Horsley}}\ and\ \bibinfo {author} {\bibfnamefont {T~G}\ \bibnamefont
  {Philbin}},\ }\bibfield  {title} {\enquote {\bibinfo {title} {Canonical
  quantization of electromagnetism in spatially dispersive media},}\ }\href
  {http://stacks.iop.org/1367-2630/16/i=1/a=013030} {\bibfield  {journal}
  {\bibinfo  {journal} {New Journal of Physics}\ }\textbf {\bibinfo {volume}
  {16}},\ \bibinfo {pages} {013030} (\bibinfo {year} {2014})}\BibitemShut
  {NoStop}%
\bibitem [{\citenamefont {Hestenes}(2002)}]{Hestenes2002}%
  \BibitemOpen
  \bibfield  {author} {\bibinfo {author} {\bibfnamefont {David}\ \bibnamefont
  {Hestenes}},\ }\enquote {\bibinfo {title} {Point groups and space groups in
  geometric algebra},}\ in\ \href {\doibase 10.1007/978-1-4612-0089-5_1} {\emph
  {\bibinfo {booktitle} {Applications of Geometric Algebra in Computer Science
  and Engineering}}},\ \bibinfo {editor} {edited by\ \bibinfo {editor}
  {\bibfnamefont {Leo}\ \bibnamefont {Dorst}}, \bibinfo {editor} {\bibfnamefont
  {Chris}\ \bibnamefont {Doran}}, \ and\ \bibinfo {editor} {\bibfnamefont
  {Joan}\ \bibnamefont {Lasenby}}}\ (\bibinfo  {publisher} {Birkh{\"a}user
  Boston},\ \bibinfo {address} {Boston, MA},\ \bibinfo {year} {2002})\ pp.\
  \bibinfo {pages} {3--34}\BibitemShut {NoStop}%
\bibitem [{\citenamefont {Nye}(1985)}]{nye1985physical}%
  \BibitemOpen
  \bibfield  {author} {\bibinfo {author} {\bibfnamefont {John~Frederick}\
  \bibnamefont {Nye}},\ }\href@noop {} {\emph {\bibinfo {title} {Physical
  properties of crystals: their representation by tensors and matrices}}}\
  (\bibinfo  {publisher} {Oxford University Press, New York},\ \bibinfo {year}
  {1985})\BibitemShut {NoStop}%
\end{thebibliography}%

\end{document}